\documentclass{article}
\usepackage{graphicx}
\usepackage{amsmath}%
\usepackage{amssymb}
\usepackage{braket}
\usepackage{tikz}
\usepackage[compat=1.1.0]{tikz-feynman}
\usetikzlibrary{arrows.meta,positioning}
\usepackage{relsize}
\usepackage[a4paper, total={7in, 10in}]{geometry}
\usepackage{enumitem}
\usepackage{dsfont}
\usepackage{authblk}
\usepackage{float} 
\usepackage{mathtools}
\usepackage[font=small]{caption}
\usepackage{physics}
\usepackage[dvipsnames]{xcolor}
\usepackage{graphicx}
\usepackage{dcolumn}
\usepackage{bm}
\usepackage[hidelinks]{hyperref}
\usepackage{amssymb}
\usepackage{url}  

\title{Interaction corrections to topological density three-point functions in two-dimensional Fermi liquids: a coadjoint orbit perspective}
\author[1]{Akshay Pal\thanks{Email: \href{akpa6771@colorado.edu}{akpa6771@colorado.edu}}}
\author[1]{Andrew Lucas}
\author[2]{Umang Mehta}

\affil[1]{Department of Physics and Center for Theory of Quantum Matter,
University of Colorado, Boulder, CO 80309, USA}

\affil[2]{School of Physics and Astronomy, University of Minnesota, Minneapolis, MN 55455, USA}
\date{}
\allowdisplaybreaks

\begin{document}

\maketitle
\begin{abstract}
  Density three-point correlations are known to probe the topology of the Fermi sea in two-dimensional noninteracting systems. Here, we study how these correlations are modified by interactions using the coadjoint-orbit effective field theory. A key advantage of the coadjoint-orbit formulation is that it provides a systematic way to incorporate generalized Landau interactions in terms of bosonized degrees of freedom, mapping fermionic loop contributions onto simpler tree-level diagrams. We show that, for a general isotropic dispersion $\epsilon(p)$, even at linear order in the generalized Landau interaction, $\mathcal{O}(\mathcal{F}^{(2,0)})$, there exists a contribution proportional to the band curvature $\epsilon''(p_F)$ that changes the nonanalytic structure of the free density three-point correlation function.
  This contribution introduces a distinct nonanalytic structure beyond that found in either the noninteracting case or an interacting Galilean-invariant system, showing that interaction effects can modify the topology-detecting density three-point correlation.
\end{abstract}
\tableofcontents
\section{Introduction}
The geometry and topology of Fermi surfaces strongly influence the properties of metallic systems. While the simplest examples involve nearly spherical Fermi surfaces, realistic materials often exhibit far richer structures, including multiply connected sheets and nontrivial topology. A classic example is copper, whose Fermi surface forms a genus-four Riemann surface~\cite{FermiSurfaceDatabase}. Such complex Fermi surfaces have been extensively characterized experimentally through angle-resolved photoemission spectroscopy (ARPES)~\cite{Borisenko} and quantum-oscillation measurements~\cite{Kwan}. Moreover, changes in Fermi-surface topology, known as Lifshitz transitions~\cite{Lifshitz}, can be induced by tuning external parameters such as pressure, doping, or carrier density in gate-controlled two-dimensional systems. Despite this progress, what remains less clear is which observables directly probe Fermi-surface topology. This question is analogous to the role of the Hall conductance in diagnosing wavefunction topology in quantum Hall systems~\cite{TKNN,BernevigBook}. In the case of Fermi surfaces, such observables have only recently been identified in noninteracting electron gases, where certain nonlinear response functions and correlation functions encode topological information about the Fermi surface~\cite{Kane2022QuantizedNC,TamKane2023AST,Tam2023PRBAST2}. Understanding how these signatures are modified, and whether any topological information survives, in interacting Fermi liquids remains an important open question.

It was recently shown that, in the ballistic limit, the Landauer conductance of a noninteracting quantum gas can be sensitive to topological changes in the Fermi sea~\cite{Kane2022QuantizedNC}. The basic idea is that an external impulse transfers particles between left- and right-moving states in a way that depends on the critical points of the band dispersion, where the group velocity vanishes. From the Morse-theoretic viewpoint, these critical points determine the Euler characteristic of the occupied region in momentum space~\cite{NashSen1988}. As a result, in the noninteracting ballistic limit, the conductance can take the topological form
\begin{equation}
\label{Eq.1}
    G = \frac{e^2}{h}\,\chi_F .
\end{equation}
This idea was further extended to higher-dimensional conductors attached to ideal point-contact terminals, where related topological features of the Fermi sea control the ballistic response~\cite{TamKane2023AST,Tam2023PRBAST2}.

However, this conductance-based topological signature is fragile. In realistic channels, electron-electron backscattering and reflections at the contacts can modify the simple ballistic result~\cite{Vanwees,Frank1998,AYacoby}. This fragility is familiar from one-dimensional interacting systems, where the conductance of a Luttinger liquid is no longer fixed solely by the noninteracting ballistic value, but is instead modified by the Luttinger parameter~\cite{KaneFisher,Egger}. Thus, while ballistic conductance provides a striking probe of Fermi-sea topology in noninteracting systems, this particular signature is not expected to remain interaction independent in the presence of interactions. Another probe that captures the topology of the Fermi sea in a $d$-dimensional noninteracting Fermi gas is the $(d+1)$-point equal-time connected density–density correlation function $\cite{Kane}$.
\begin{align}
\label{Eq.2}
    S_{d+1}(q_{1},...q_{d})&=\int \frac{d^{d}q_{d+1}}{(2\pi)^{d}}\langle \rho_{q_1}....\rho_{q_{d+1}}\rangle_{c} = \frac{V_{q_{\alpha}}}{(2\pi)^{d}}\chi_{F}\xRightarrow{d=2}S_{3}(\mathbf{q_1},\mathbf{q_2})=\frac{|\mathbf{q_1}\times \mathbf{q_2}|}{(2\pi)^2}\chi_F
\end{align}
where $\chi_{F}$ is the Euler characteristic of the  $d$-dimensional Fermi sea. $V_{q_{\alpha}}$ is the  $d$-dimensional volume of the 
parallelepiped spanned by the $d$-basis vectors $q_{\alpha}$. This prediction was recently confirmed experimentally using single-atom-resolved
imaging of a two-dimensional gas of $^{6}\mathrm{Li}$ atoms
\cite{Daix2025_FermiSea}, which has motivated theoretical work on how
interactions modify Eq.~\eqref{Eq.2}. In a long-wavelength collinear limit, the
singularity survives with a coefficient renormalized by Landau parameters
\cite{Tam2026Singular}, and for a weakly interacting gas with a contact
interaction the total-density correlator has been obtained exactly to first
order in the interaction strength \cite{Kane2026ThreePoint}. 

Here we approach the problem from the effective theory of a Fermi liquid,
treating the Landau interaction and the quasiparticle dispersion as independent
inputs. Using the coadjoint-orbit formulation, we develop a diagrammatic expansion for $S_3$. At first order in the Landau interaction, we isolate a new nonanalytic contribution proportional to the band curvature $\epsilon''(p_F)$, whose momentum dependence is distinct from that of the noninteracting result. It is smooth at the parallel configuration, $\Delta\varphi=0$, where
$|\mathbf{q}_1\times\mathbf{q}_2|$ vanishes, but develops a nonanalytic cusp at the
antiparallel configuration, $\Delta\varphi=\pi$. It therefore constitutes a distinct
structure from the collinear singularity whose coefficient is renormalized by the
Landau parameters~\cite{Tam2026Singular,Kane2026ThreePoint}. Band curvature
enters through the dimensionless ratio $p_F\epsilon''(p_F)/v_F$, which is fixed
for a Galilean-invariant gas but is a free parameter for a general isotropic dispersion,
so that the nonanalytic structure of $S_3$ is not determined by the Landau parameters alone.

This paper is organized as follows. Sec.~\ref{Sec 2} reviews the one-dimensional
case, where the low-energy theory is a Luttinger liquid and conformal invariance
organizes how interactions enter correlation functions. There, the density
two-point function retains its nonanalytic $|q|$ form, but its coefficient is
rescaled by the Luttinger parameter, so that it no longer probes the topology of the Fermi sea in an interaction-independent way. The two-point function of the
stress tensor does, however, retain this character, since it is controlled by the
central charge. Sec.~\ref{Sec 3} turns to two dimensions. Using the coadjoint-orbit method, we
set up a diagrammatic expansion of the three-point density correlation function,
recover the noninteracting result, and organize the first-order corrections in
the Landau interaction. We identify a contribution proportional to the band
curvature $\epsilon''(p_F)$, which can be isolated since no other diagram at this
order depends on it. This term produces a nonanalytic structure whose angular
dependence is distinct from that of the noninteracting result, demonstrating that
interactions can generate structures beyond a renormalization of the
noninteracting form. We further show that this contribution is consistent with
the symmetry constraints. Finally, in Sec.~\ref{Sec 4}, we discuss the implications
of these results for the topology-probing character of density correlations in
interacting Fermi liquids.  

\section{Topological correlators in 1D}
\label{Sec 2}
In one spatial dimension, the Fermi sea consists of a collection of disjoint
intervals in momentum space, whose boundaries form a discrete set of Fermi
points carrying right- or left-moving chirality. Throughout this section we
consider spinless fermions in a single band, so that each interval of the Fermi
sea carries one pair of Fermi points and, at low energies, a single gapless
mode. For free fermions, Ref.~\cite{Kane} showed that this topology is encoded
directly in the long-wavelength equal-time density correlations,
\begin{equation}
  \label{eq:S2free}
  S_2^{\mathrm{free}}(q)
  =
  \langle \rho_q\rho_{-q}\rangle
  =
  \frac{|q|}{2\pi}\,\chi_F ,
\end{equation}
where $\chi_F$ is the Euler characteristic of the Fermi sea, equal in one
dimension to the number of its connected components. Once interactions are included, the situation becomes more subtle. The low-energy theory remains
gapless and the set of Fermi points is unchanged, yet correlation functions
generally acquire a dependence on the marginal couplings of the
Luttinger-liquid fixed point. To see this explicitly, we begin with a single connected Fermi interval, whose
Fermi surface consists of two points corresponding to right- and left-moving
fermions. At low energies the Hamiltonian may be written in terms of the
associated chiral densities as
\begin{equation}
  \label{eq:Hchiral}
  H = \int\!dx\,\Big[
  (\pi v_F+g_4)\big(\rho_R^2+\rho_L^2\big)
  +2g_2\,\rho_R\rho_L
  \Big],
\end{equation}
where $v_F$ is the Fermi velocity of a parity-invariant dispersion relation.
The couplings $g_4$ and $g_2$ describe intrabranch and interbranch scattering,
respectively, and constitute the two marginal interactions of the theory. The chiral densities satisfy a Kac-Moody algebra whose structure is
independent of the interaction strengths~\cite{Giamarchi2003,Fradkin2013}. The
strongly constrained kinematics of one dimension further allow the low-energy
excitations to be recast in terms of a compact boson $\phi$ and its dual
$\vartheta$, a correspondence known as bosonization. In this representation
the physical and chiral densities are
\begin{equation}
  \label{eq:bosonization}
  \rho=\frac{1}{\sqrt{\pi}}\,\partial_x\phi ,
  \qquad
  \rho_5=\frac{1}{\sqrt{\pi}}\,\partial_x\vartheta ,
  \qquad
  \rho_{R/L}=\tfrac{1}{2}\left(\rho\pm\rho_5\right),
\end{equation}
with $\Pi=\partial_x\vartheta$ canonically conjugate to $\phi$, while the
Hamiltonian takes the standard Gaussian form
\begin{equation}
  \label{eq:Hgauss}
  H=
  \frac{v}{2}\int\!dx\,
  \left[
  K(\partial_x\phi)^2
  +\frac{1}{K}(\partial_x\vartheta)^2
  \right].
\end{equation}
The interaction-renormalized velocity $v$ and the Luttinger parameter $K$ are
given by
\begin{equation}
  \label{eq:vK}
  v=
  \left[
  \left(v_F+\frac{g_4}{\pi}\right)^2
  -
  \left(\frac{g_2}{\pi}\right)^2
  \right]^{1/2},
  \qquad
  K=
  \left(
  \frac{\pi v_F+g_4+g_2}
       {\pi v_F+g_4-g_2}
  \right)^{1/2} .
\end{equation}
We now turn to the effect of interactions on the equal-time density two-point
function discussed above, which becomes
\begin{equation}
\label{eq:int_Luttinger_density_2pt}
  S_2^{\mathrm{int}}(q)
  =
  \langle\rho_q\rho_{-q}\rangle
  =
  \frac{1}{K}\,\frac{|q|}{2\pi}
  =
  \frac{1}{K}\,S_2^{\mathrm{free}}(q) .
\end{equation}
Interactions therefore preserve the characteristic nonanalytic $|q|$ dependence
of the free-fermion result, but renormalize its coefficient by the Luttinger
parameter\footnote{Our convention for the Luttinger parameter $K$ is the
reciprocal of that adopted in Ref.~\cite{Kane}.}, so that the result no longer
retains the interaction-independent topological character of
Eq.~\eqref{eq:S2free}.
\par
This raises a broader question: how does interaction dependence appear in
correlation functions at the Luttinger-liquid fixed point, and under what
circumstances can it disappear? We will not attempt a general classification,
but the conformal structure of the theory provides a useful framework for
organizing the discussion. For a single Fermi interval, the Luttinger-liquid
fixed point is a $c=1$ conformal field theory of a compact free boson, and the
Luttinger parameter $K$ labels a continuous family of interacting conformal
fixed points. Interactions thus preserve the conformal structure of the theory
and the corresponding constraints on correlation functions, while quantities
such as scaling dimensions and correlation-function amplitudes can vary
continuously with $K$.
\par
A simple illustration is provided by the primary operators
$\partial\phi\equiv\partial_z\phi$ and
$\bar\partial\phi\equiv\partial_{\bar z}\phi$, with $z=x+iy$, which are holomorphic and
antiholomorphic primaries with conformal weights $(1,0)$ and $(0,1)$,
respectively. These weights are independent of the Luttinger parameter $K$, so
their scaling dimensions remain fixed along the Luttinger-liquid fixed line.
Nevertheless, when the bosonic field is normalized according to the microscopic
density convention in Eq.~\eqref{eq:bosonization}, the corresponding two-point
functions retain an explicit dependence on $K$:
\begin{equation}
  \label{eq:dphidphi}
  \big\langle \partial\phi(z)\,\partial\phi(w)\big\rangle
  =
  -\frac{1}{4\pi K}\frac{1}{(z-w)^2},
  \qquad
  \big\langle \bar\partial\phi(\bar z)\,\bar\partial\phi(\bar w)\big\rangle
  =
  -\frac{1}{4\pi K}\frac{1}{(\bar z-\bar w)^2}.
\end{equation}
Interactions therefore need not change the conformal dimension of a primary
operator in order to modify its correlation functions; the interaction
dependence may instead enter through the overall normalization. By contrast, the vertex primary operators provide a complementary example in
which the interaction dependence enters directly through the conformal
dimensions. The operators
\begin{equation}
  \mathcal{V}_{m,n}
  =
  :\exp\left[
  i\sqrt{\pi}\left(m\phi+n\vartheta\right)
  \right]:,
  \qquad
  m,n\in\mathbb{Z},
\end{equation}
have holomorphic and antiholomorphic conformal weights
\begin{equation}
  h_{m,n}
  =
  \frac{1}{8}
  \left(
  \frac{m}{\sqrt{K}}+n\sqrt{K}
  \right)^2,
  \qquad
  \bar h_{m,n}
  =
  \frac{1}{8}
  \left(
  \frac{m}{\sqrt{K}}-n\sqrt{K}
  \right)^2.
\end{equation}
Thus, unlike the derivative primaries, the conformal dimensions of the vertex
operators vary continuously with the Luttinger parameter $K$. This
interaction dependence is directly reflected in their two-point functions,
whose coordinate dependence is fixed by conformal symmetry:
\begin{equation}
  \left\langle
  \mathcal{V}_{m,n}(z,\bar z)\,
  \mathcal{V}_{-m,-n}(w,\bar w)
  \right\rangle
  \sim
  \frac{1}
  {(z-w)^{2h_{m,n}}
   (\bar z-\bar w)^{2\bar h_{m,n}}}.
\end{equation}

Interactions therefore modify not only the normalization of correlation
functions but, for vertex operators, the power-law exponents themselves.
Since the primary operators of the Luttinger-liquid CFT acquire $K$ dependence
either through their normalization or through their conformal dimensions, one
may expect higher-point functions built from them to depend on $K$ as well.
This provides only a qualitative guide, since particular correlators need not follow this expectation: the
dependence may cancel, or the correlator may vanish identically. The stress tensor provides a particularly instructive counterexample. Although
it is not a primary operator, its correlation functions are fixed by the
Virasoro algebra and are controlled by the central charge rather than by the
Luttinger parameter. With the normalization adopted above, the holomorphic
stress tensor takes the form
\begin{equation}
    T(z)
    =
    -2\pi K
    :\!\bigl(\partial\phi(z)\bigr)^2\!:\, .
\end{equation}
Using the two-point function of $\partial\phi$ given above, one immediately
finds
\begin{equation}
    \left\langle
    T(z)T(w)
    \right\rangle
    =
    \frac{1}{2(z-w)^4},
\end{equation}
which identifies the theory as having central charge $c=1$, independent of $K$. Thus, although the constituent
fields carry explicit $K$ dependence, this dependence cancels in the
stress-tensor correlator, whose coefficient is fixed by the central charge
alone. 

This observation has a direct topological interpretation when the Fermi sea
contains several disconnected intervals. For a Fermi sea consisting of $N$
intervals the low-energy theory contains $N$ gapless compact bosonic modes,
provided that interactions do not gap any linear combination of them. Couplings
between different intervals mix these modes but, being exactly marginal, leave
the total central charge unchanged, and central charges are additive, so that
\begin{equation}
  \label{eq:ctot}
  c_{\mathrm{tot}}=N=\chi_F ,
\end{equation}
and the total stress tensor therefore satisfies
\begin{equation}
  \label{eq:TTtot}
  \big\langle
  T_{\mathrm{tot}}(z)\,T_{\mathrm{tot}}(w)
  \big\rangle
  =
  \frac{\chi_F}{2(z-w)^4} .
\end{equation}
The Euler characteristic is thus encoded in an interaction-independent
low-energy correlation function, even though the ordinary density two-point
function is renormalized by interactions. 

\section{Weakly interacting 2D Fermi liquids}
\label{Sec 3}
In two spatial dimensions, Ref.~\cite{Kane} demonstrated that the equal-time density three-point function directly probes the topology of the free Fermi sea. Specifically,
\begin{equation}\label{eq:density_3pt_free}
S^\mathrm{free}_3(\mathbf{k},\mathbf{k}_1)
= \frac{|\mathbf{k}\times\mathbf{k}_1|}{(2\pi)^2}\chi_F,
\end{equation}
where the Euler characteristic of the two-dimensional Fermi sea is
\begin{equation}
\chi_F = N_e-N_h,
\end{equation}
with $N_e$ and $N_h$ denoting the numbers of electron-like and hole-like Fermi surfaces, respectively. It is therefore natural to ask how this relation is modified in the presence of interactions. In particular, we ask whether interactions merely renormalize the coefficient of the noninteracting structure, or whether they can generate additional momentum-dependent structures, thereby modifying the topology-probing character of the density three-point function. While conformal field theory provides a useful framework for understanding interaction effects in one dimension, the higher-dimensional problem is more involved. For $d>1$, the low-energy theory of a Fermi surface is scale invariant but not conformally invariant, in part because the Fermi energy $E_F$ provides an intrinsic scale that remains fixed under the low-energy scaling transformation \cite{Shankar1994,Polchinski1992}. Rather than attempting a fully nonperturbative treatment of the interacting problem, we therefore adopt the more modest goal of computing perturbative corrections to Eq.~\eqref{eq:density_3pt_free} generated by Landau interactions. At the level of linear response, these interactions are parametrized by the
usual Landau parameters $F_l$, which constitute the marginal couplings of the
Fermi-liquid fixed point and play a role analogous to the $g_2$ and $g_4$
couplings in one dimension. The three-point
function, however, is a nonlinear response, and is sensitive both to further
couplings in the effective Hamiltonian and to the curvature of the quasiparticle
dispersion~\cite{DelacretazDuttaChowdhuryMehta2025}.

For this purpose, the coadjoint-orbit formulation provides a particularly natural framework \cite{Umang,Mehta2023}. It gives a bosonized description of the Fermi liquid in which the fundamental degrees of freedom are deformations of the Fermi surface, while the nonlinear algebra governing these fluctuations is incorporated directly into the effective action. This feature is especially important for the density three-point function, whose leading nontrivial contribution is intrinsically sensitive to nonlinear Fermi-surface dynamics and therefore cannot be captured by a purely Gaussian theory. At the same time, bosonization reorganizes the conventional fermionic perturbative expansion: contributions that would arise from fermion-loop diagrams are encoded in nonlinear vertices of the bosonic effective theory and can consequently appear at the tree level. The coadjoint-orbit formalism thus provides both a systematic description of the nonlinear kinematics of Fermi-surface fluctuations and an efficient perturbative framework for determining how Landau interactions modify the topological three-point function in Eq.~\eqref{eq:density_3pt_free}.

\subsection{Coadjoint orbit action}

We adopt the coadjoint-orbit effective field theory for interacting  fermions in the long-wavelength regime, where the relevant particle-hole excitations are confined to the vicinity of the Fermi surface and carry momenta much smaller than the local Fermi momentum. Consequently, these low-energy excitations probe only small and smooth deformations of the Fermi surface within each local patch of phase space. Rather than keeping track of the microscopic fermionic degrees of freedom individually, it is therefore natural to describe the low-energy dynamics directly in terms of these Fermi-surface deformations. Such deformations can be viewed as arising from the action of canonical transformations on phase space. Since canonical transformations preserve phase-space volume, they deform the shape of the Fermi surface without changing the volume it encloses, consistent with Luttinger's theorem, which fixes the Fermi volume at a given particle density. Let $\mathfrak{g}$ denote the Lie algebra associated with the Lie group $G$ of canonical transformations \cite{Umang,Mehta2023}. In the long-wavelength limit relevant to our effective description, this Lie algebra reduces to the familiar Poisson algebra. We introduce a bosonic field $\phi(\mathbf{x},\theta)\in\mathfrak{g}$, which parametrizes the canonical transformation $U=\exp(-\phi)$. Physically, $\phi(\mathbf{x},\theta)$ describes a deformation of the Fermi surface at spatial position $\mathbf{x}$ and angular coordinate $\theta$ along the Fermi surface in momentum space. Starting from a reference distribution $f_0$, the deformed phase-space distribution is obtained through the coadjoint action,
\begin{align} \label{eqn: coadjoint orbit action} f_{\phi}=\text{Ad}^{*}_{U=\exp(-\phi)}f_{0}=Uf_{0}U^{-1}=f_{0}-\{\phi,f_{0}\}+\frac{1}{2!}\{ \phi ,\{\phi,f_{0}\}\}+\cdots 
\end{align}
where $f_0$ denotes the equilibrium Fermi distribution and $\{A,B\}=\nabla_{x}A\cdot\nabla_{p}B-\nabla_{p}A\cdot\nabla_{x}B$ defines the Poisson bracket. The phase-space distribution $f(\mathbf{x},\mathbf{p})$ can be viewed as an element of the dual space $\mathfrak{g}^{*}$ associated with the Lie algebra $\mathfrak{g}$. In other words, $f$ acts as a linear functional on elements $F\in\mathfrak{g}$, assigning to each $F$ a number through the pairing
\begin{align}
    \label{eqn:dual_pairing}
    f[F]
    \equiv \langle f,F\rangle
    \equiv
    \int \frac{d^2 x\,d^2 p}{(2\pi)^2}\,
    f(\mathbf{x},\mathbf{p})F(\mathbf{x},\mathbf{p}).
\end{align}
Our next step is to construct a Hamiltonian functional $H[f]$ using effective field theory by writing an expansion in the nonlinearities of $f$:
\begin{align}
    \label{eqn: Effective action}
    H[f]&=\int_{\textbf{x},\textbf{p}}\epsilon(\textbf{p})f_{\phi}(\textbf{x,p})+\frac{1}{2}\int_{\textbf{x,p,p}'}\mathcal{F}^{(2,0)}(\textbf{p},\textbf{p}')\delta f_{\phi}(\textbf{x},\textbf{p})\delta f_{\phi}(\textbf{x},\textbf{p}')\notag\\&\quad+\frac{1}{2}\int_{\textbf{x,p,p}'}\mathcal{F}^{(2,1)}(\mathbf{p,\mathbf{p'}})\cdot \Big(\frac{\nabla_{x}}{p_{F}}\delta f_{\phi}(\mathbf{x},\mathbf{p})\Big)\delta f_{\phi}(\mathbf{x},\mathbf{p'})\notag\\&\qquad+
    \frac{1}{3}\int_{\textbf{x,p,p}'}\mathcal{F}^{(3,0)}(\mathbf{p},\mathbf{p'},\mathbf{p''})\delta f_{\phi}(\mathbf{x},\mathbf{p})\delta f_{\phi}(\mathbf{x},\mathbf{p'})\delta f_{\phi}(\mathbf{x},\mathbf{p''})\cdots
\end{align}
where $\epsilon(\textbf{p})$ is the dispersion relation, $\delta f(\textbf{x},\textbf{p})\equiv f(\textbf{x},\textbf{p})-f_{0}(\textbf{p})$ is the deviation from the equilibrium distribution and 
$\mathcal{F}^{(n,m)}$ denotes the generalized Landau interaction function where the $n$-index labels the nonlinearity of interactions and $m$-index labels the number of x-derivatives in that coupling. The effective action is obtained from the Hamiltonian formulation through a Legendre transformation. Unlike in an ordinary Hamiltonian system, there is no natural choice of canonical coordinates and conjugate momenta for the Fermi-surface distribution. Instead, the coadjoint orbit carries a natural Kirillov--Kostant--Souriau (KKS) symplectic structure \cite{ArnoldKhesin1998}, which plays the role of the canonical phase-space structure and gives rise to the Wess--Zumino--Witten term. The resulting effective action is
\begin{align}
    \label{eqn:Effective_Action}
    S
    &= S_{\mathrm{WZW}} + S_H \notag\\
    &= \int dt\, \langle f_0,U^{-1}\partial_t U\rangle
    - \int dt\, H[f_{\phi}] .
\end{align}

We now use the coadjoint orbit effective theory to study density correlation functions arising from Fermi-surface fluctuations \cite{DelacretazDuttaChowdhuryMehta2025}. The density fluctuation is related to the deformation of the phase-space distribution through
\begin{align}
    \delta \rho(\mathbf{x})
    =
    \int_{\mathbf{p}}
    \delta f_{\phi}(\mathbf{x},\mathbf{p}),
\end{align}
where $\int_{\mathbf{p}}\equiv\int d^2 p/(2\pi)^2$. In momentum space, the equal-time three-point density correlation function is defined as
\begin{align}
    S_3(\mathbf{q},\mathbf{q}')
    &=
    \int \frac{d\omega}{2\pi}
    \frac{d\omega'}{2\pi}
    \left\langle
    \delta\rho(\omega,\mathbf{q})\,
    \delta\rho(\omega',\mathbf{q}')\,
    \delta\rho(-\omega-\omega',-\mathbf{q}-\mathbf{q}')
    \right\rangle ,
\end{align}
where translational invariance enforces conservation of frequency and momentum. Unlike the two-point function, the three-point correlation function is intrinsically sensitive to the nonlinear structure of Fermi-surface fluctuations, making it a natural observable within the coadjoint-orbit effective theory.

\subsection{Noninteracting equal-time three-point density correlations}
The effective action admits two complementary perturbative expansions. The first is the usual effective-field-theory expansion in generalized Landau interaction functions $\mathcal{F}^{(n,m)}$, which characterize interactions among quasiparticles and gradient corrections associated with spatial variations of the distribution function. Schematically, the action can be organized as
\begin{align}
S = S_{\mathrm{free}} +S_{\mathcal{F}^{(2,0)}} +S_{\mathcal{F}^{(2,1)}} +S_{\mathcal{F}^{(3,0)}} +\cdots . \end{align} 
Each term in the effective-field-theory expansion can itself be expanded in powers of the Fermi-surface deformation field $\phi(\mathbf{x},\theta)$. Denoting the contribution of order $\phi^r$ by a superscript $[r]$, we write schematically
\begin{align}
    S_X
    =
    S_X^{[2]}
    +S_X^{[3]}
    +S_X^{[4]}
    +\cdots,
    \qquad
    X\in
    \left\{
    \mathrm{Free},
    \mathcal{F}^{(2,0)},
    \mathcal{F}^{(2,1)},
    \mathcal{F}^{(3,0)},
    \ldots
    \right\}.
\end{align}
In this work, we restrict ourselves to the simplest form of the Landau interaction, assuming that $\mathcal{F}^{(2,0)}$ depends only on the angular directions of the momenta, such that $\partial_{|\mathbf{p}|}\mathcal{F}^{(2,0)}(\mathbf{p},\mathbf{p}')=0$ and neglect all higher-order quasiparticle interactions by setting $\mathcal{F}^{(n,m)}=0$ for $n\geq 3$ or $m>0$.

For the present subsection, we make the additional restriction of considering only the free part of the effective action, since our immediate goal is to evaluate the noninteracting contribution to the equal-time three-point density correlation function. Expanding the free action to cubic order in the Fermi-surface deformation field $\phi$, we write

\begin{align}
    S_{\mathrm{free}}
    &=
    S_{\mathrm{G}}^{[2]}
    +S_{\mathrm{WZW},\text{free}}^{[3]}
    +S_{H,\text{free}}^{[3]}
    \notag\\
    &=
    -\frac{p_F}{2}
    \int_{t,\mathbf{x},\theta}
    \nabla_n\phi
    \left(
    \dot{\phi}
    -v_F\nabla_n\phi
    \right)
    \notag\\
    &\quad
    -\frac{1}{3!}
    \int_{t,\mathbf{x},\theta}
    \nabla_n\phi
    \left[
    \nabla_s\phi\,\partial_\theta\dot{\phi}
    -
    \nabla_s\dot{\phi}\,\partial_\theta\phi
    \right]
    \notag\\
    &\quad
    -\frac{1}{3!}
    \int_{t,\mathbf{x},\theta}
    \left(
    \frac{v_F}{2}
    +p_F\epsilon''
    \right)
    \left(\nabla_n\phi\right)^3 .
\end{align}
Here, $S_{\mathrm{G}}^{[2]}$ denotes the quadratic Gaussian action, which determines the propagator of the bosonic Fermi-surface fluctuations. The two cubic terms arise respectively from the Wess-Zumino-Witten sector and from the nonlinear expansion of the Hamiltonian. We use $\dot{\phi}\equiv\partial_t\phi$ to denote the time derivative of $\phi$. The vector $\mathbf{n}_\theta$ is the outward unit normal to the Fermi surface at angular coordinate $\theta$, while $\mathbf{s}_\theta\equiv\partial_\theta\mathbf{n}_\theta$ denotes the corresponding tangent vector. We use the shorthand $\nabla_n\phi\equiv\mathbf{n}_\theta\cdot\nabla_{\mathbf{x}}\phi$ and $\nabla_s\phi\equiv\mathbf{s}_\theta\cdot\nabla_{\mathbf{x}}\phi$. Finally, $v_F=\epsilon'(p_F)$ denotes the Fermi velocity, while $\epsilon''\equiv\epsilon''(p_F)$ characterizes the curvature of the quasiparticle dispersion at the Fermi surface. Another important point is that the density operator is itself a nonlinear functional of the deformation field $\phi$. It therefore admits a perturbative expansion of the form
\begin{align}
    \delta \rho
    =
    \delta \rho^{[1]}
    +
    \delta \rho^{[2]}
    + \cdots .
\end{align}
By Wick’s theorem, the Gaussian theory with three linear-density insertions does not generate a connected three-point correlator.
The leading non-vanishing contribution therefore arises from the cubic Wess--Zumino--Witten vertex $S_{\mathrm{WZW},\text{free}}^{[3]}$, the cubic Hamiltonian vertex $S_{H,\text{free}}^{[3]}$, and the quadratic term $\delta\rho^{[2]}$ in the density operator. Accordingly, the connected three-point function can be decomposed as
\begin{align}
    \label{eq:S3_decomposition}
    \left\langle
    \delta \rho\,\delta \rho\,\delta \rho
    \right\rangle_c
    &=
    \left\langle
    \delta \rho\,\delta \rho\,\delta \rho
    \right\rangle_c^{(\mathrm{WZW})}
    +
    \left\langle
    \delta \rho\,\delta \rho\,\delta \rho
    \right\rangle_c^{(H)}
    +
    \left\langle
    \delta \rho\,\delta \rho\,\delta \rho
    \right\rangle_c^{(\rho^{[2]})} .
\end{align}
Diagrammatically, this decomposition is represented as
\begin{equation}
\label{eq:S3_diagrams}
\left\langle
\delta \rho\,\delta \rho\,\delta \rho
\right\rangle_c
=
\vcenter{\hbox{
\begin{tikzpicture}[scale=0.9, line cap=round, line join=round]

\tikzset{
  prop/.style={thick},
  dprop/.style={thick, dashed},
  wzwvertex/.style={circle, fill=black, inner sep=1.8pt},
  hvertex/.style={rectangle, draw=red!80!black, fill=red!75, inner sep=2.5pt},
  ins/.style={circle, draw=black, line width=0.9pt, inner sep=1.2pt},
  lab/.style={font=\footnotesize}
}

\coordinate (V1) at (-3.8,0.8);
\node[wzwvertex] at (V1) {};

\draw[prop]  (V1) -- ++(0,1.9);
\draw[prop]  (V1) -- ++(-1.4,-1.6);
\draw[dprop] (V1) -- ++( 1.4,-1.6);

\node[lab] at (-3.8,-1.55) {$S_{\mathrm{WZW},\text{free}}^{[3]}$};

\node at (-1.15,0.8) {$+$};

\coordinate (V2) at (1.3,0.8);
\node[hvertex] at (V2) {};

\draw[prop] (V2) -- ++(0,1.9);
\draw[prop] (V2) -- ++(-1.4,-1.6);
\draw[prop] (V2) -- ++( 1.4,-1.6);

\node[lab] at (1.3,-1.55) {$S_{H,\text{free}}^{[3]}$};

\node at (4.0,0.8) {$+$};

\coordinate (V3) at (7.2,0.8);
\node[wzwvertex] at (V3) {};

\draw[prop]  (V3) -- ++(0,0.5);
\draw[prop]  (V3) -- ++(-1.4,-1.6);
\draw[dprop] (V3) -- ++( 1.4,-1.6);

\node[lab] at (7.2,-1.55) {$\delta\rho^{[2]}$};

\end{tikzpicture}
}}
\end{equation}
The equal-time three-point density correlation function is obtained by integrating the two external frequencies. The details of the Feynman-diagram evaluation and the frequency integrations are deferred to Appendix \ref{app: Derivation_NI_correlations}. After performing these integrals, one finds that the contributions arising from the cubic Wess--Zumino--Witten vertex $S_{\mathrm{WZW},\text{free}}^{[3]}$ and the cubic Hamiltonian vertex $S_{H,\text{free}}^{[3]}$ vanish. Notably, $S_{H,\text{free}}^{[3]}$  contains $\epsilon''(p_{F})$, but its contribution vanishes after the frequency integrations. The only surviving contribution originates from the quadratic term $\delta\rho^{[2]}$ in the density operator. For a single simply connected Fermi sea, for which $\chi_F=1$, the resulting equal-time three-point density correlation function is
\begin{equation}
    \label{eq:density_3pt_free_for_sphere}
    S^\mathrm{free}_3(\mathbf{k},\mathbf{k}_1)
    =
    \frac{|\mathbf{k}\times\mathbf{k}_1|}{(2\pi)^2}.
\end{equation}
This reproduces the universal noninteracting result of Tam and Kane~\cite{Kane}, obtained from a geometric triangulation of the Fermi sea.

\subsection{Interaction correction from band curvature to the density three-point function}
\label{sec:eps2correction}

We now show that, even at leading order in the Landau interaction, the density three-point function acquires a second nonanalytic structure with an angular dependence distinct from that of the noninteracting result. Its coefficient depends explicitly on the band curvature and is therefore not fixed by the Landau parameters alone. We do not compute the complete first-order correction; instead, we isolate a single independent contribution with different momentum dependence.

To exhibit this, we retain terms
of order $\mathcal{O}(\mathcal{F}^{(2,0)})$ in the effective action. Expanding them in powers of the Fermi-surface deformation
field $\phi$, we obtain
\begin{align}
    S_{\mathcal{F}^{(2,0)}}
    &=
    S_{H,\mathcal{F}^{(2,0)}}^{[2]}
    +
    S_{H,\mathcal{F}^{(2,0)}}^{[3]}
    +\cdots
    \notag\\
    &=
    -\frac{p_F}{2}
    \int_{t,\mathbf{x},\theta,\theta'}
    v_F\,
    \mathcal{F}^{(2,0)}(\theta,\theta')\,
    \nabla_n\phi\,
    \nabla_{n'}\phi'
    \notag\\
    &\quad
    -\frac{1}{2}
    \int_{t,\mathbf{x},\theta,\theta'}
    v_F\,
    \mathcal{F}^{(2,0)}(\theta,\theta')
    \Big[
        \big(
            \nabla_s\nabla_n\phi\,\partial_\theta\phi
            -
            \partial_\theta\nabla_n\phi\,\nabla_s\phi
        \big)
        \nabla_{n'}\phi'
        +(\theta\leftrightarrow\theta')
    \Big]
    +\cdots
\end{align}
where \begin{equation}
    \mathcal{F}^{(2,0)}(\theta,\theta')=\frac{p_F}{v_F}\mathcal{F}^{(2,0)}(p_F\hat{n}_{\theta},p_F\hat{n}_{\theta'}) 
\end{equation}is the dimensionless Landau interaction.
For a rotationally invariant Fermi liquid, the Landau interaction can be decomposed into angular harmonics as \begin{equation}
    \mathcal{F}^{(2,0)}(\theta,\theta')
=
F_{0}
+
2\displaystyle\sum_{l=1}^{\infty}
F_l\cos\!\left[l(\theta-\theta')\right]
\end{equation}
where $F_l$ denote the corresponding Landau parameters. 

The correction to the density three-point function at first order in the Landau interaction receives contributions from both the quadratic and cubic pieces of the interaction-dependent action. Schematically,
\begin{align}
    \left\langle
    \delta\rho\,\delta\rho\,\delta\rho
    \right\rangle_{\mathcal{O}(\mathcal{F})}
    &=
    \left\langle
    \delta\rho\,\delta\rho\,\delta\rho
    \right\rangle_{S^{[3]}_{H,\mathcal{F}^{(2,0)}}}
    +
    \left\langle
    \delta\rho\,\delta\rho\,\delta\rho
    \right\rangle_{
        S^{[2]}_{H,\mathcal{F}^{(2,0)}}
        \times
        S^{[3]}_{\mathrm{WZW,free}}
    }
    +
    \left\langle
    \delta\rho\,\delta\rho\,\delta\rho
    \right\rangle_{
        S^{[2]}_{H,\mathcal{F}^{(2,0)}}
        \times
        S^{[3]}_{H,\mathrm{free}}
    }
    +
    \left\langle
    \delta\rho\,\delta\rho\,\delta\rho
    \right\rangle_{
        S^{[2]}_{H,\mathcal{F}^{(2,0)}}
        \times
        \delta\rho^{[2]}
    } .
\end{align}
 These contributions have a simple diagrammatic origin.
 The cubic interaction term $S^{[3]}_{H,\mathcal{F}^{(2,0)}}$ generates a new three-point vertex and therefore contributes directly to the density three-point function. The quadratic interaction term $S^{[2]}_{H,\mathcal{F}^{(2,0)}}$, on the other hand, does not generate a three-point vertex directly, but instead modifies the bosonic propagator at linear order in the interaction. Consequently, each of the three-point diagrams already present in the free theory acquires a first-order correction through an insertion of $S^{[2]}_{H,\mathcal{F}^{(2,0)}}$ into one of its propagator lines. This gives rise to the mixed contributions involving the free cubic Wess-Zumino-Witten vertex $S^{[3]}_{\mathrm{WZW,free}}$, the free cubic Hamiltonian vertex $S^{[3]}_{H,\mathrm{free}}$, and the nonlinear density insertion $\delta\rho^{[2]}$. Physically, these mixed terms describe the interaction-induced dressing of the propagation of Fermi-surface fluctuations, while $S^{[3]}_{H,\mathcal{F}^{(2,0)}}$ represents a genuinely new nonlinear vertex generated by the Landau interaction.


\begin{equation}
\label{eq:O_F_diagrams}
\left\langle
\delta\rho\,\delta\rho\,\delta\rho
\right\rangle_{\mathcal O(\mathcal F^{(2,0)})}
=
\vcenter{\hbox{
\begin{tikzpicture}[line cap=round,line join=round,scale=0.95]

\tikzset{
    prop/.style={line width=1.15pt},
    dprop/.style={line width=1.05pt,dashed,dash pattern=on 4pt off 4pt},
    wzwvertex/.style={circle,fill=black,inner sep=1.6pt},
    hvertex/.style={rectangle,draw=red!80!black,fill=red!75,inner sep=2.4pt},
    ins/.style={circle,draw=black,line width=0.9pt,inner sep=1.3pt,fill=white},
    lab/.style={font=\small},
    biglab/.style={font=\large},
    brace/.style={decorate,decoration={brace,amplitude=6pt}}
}


\coordinate (A) at (-1.4,0.0);
\node[hvertex] at (A) {};
\draw[prop]  (A) -- ++(0,1.15);
\draw[prop]  (A) -- ++(-1.10,-1.65);
\draw[dprop] (A) -- ++( 1.10,-1.65);

\node[
    circle,
    draw=black,
    line width=0.8pt,
    inner sep=0.5pt,
    font=\scriptsize
] at (A) {$\times$};

\node[lab] at ($(A)+(0.55,0.15)$) {$\mathcal F$};

\node[biglab] at (0.35,-0.25) {$+$};

\coordinate (B) at (2.2,0.0);
\node[wzwvertex] at (B) {};
\draw[prop]  (B) -- ++(0,1.15);
\draw[prop]  (B) -- ++(-1.10,-1.65);
\draw[dprop] (B) -- ++( 1.10,-1.65);

\coordinate (XB) at ($(B)!0.52!($(B)+(-1.10,-1.65)$)$);
\node[ins] at (XB) {$\times$};
\node[lab] at ($(XB)+(0.2,-0.4)$) {$\mathcal F$};

\node[biglab] at (4.35,-0.25) {$+$};

\coordinate (C) at (6.1,0.0);
\node[wzwvertex] at (C) {};
\draw[prop]  (C) -- ++(0,1.15);
\draw[prop]  (C) -- ++(-1.10,-1.65);
\draw[dprop] (C) -- ++( 1.10,-1.65);

\coordinate (XC) at ($(C)!0.52!($(C)+(1.10,-1.65)$)$);
\node[ins] at (XC) {$\times$};
\node[lab] at ($(XC)+(0.1,-0.52)$) {$\mathcal F$};

\node[lab] at (-1.8,-2.3) {$S_{H,\mathcal{F}^{(2,0)}}^{[3]}\ \text{part}$};
\draw[brace,decoration={brace,mirror,amplitude=6pt}]
(1.0,-1.8) -- (7.3,-1.8)
node[midway,yshift=-0.5cm] {$S_{H,\mathcal{F}^{(2,0)}}^{[2]}\times S^{[3]}_{\mathrm{WZW},\text{Free}}\ \text{part}$};


\node[biglab] at (-4.3,-4.0) {$+$};

\coordinate (D) at (-1.8,-4.0);
\node[hvertex] at (D) {};
\draw[prop] (D) -- ++(0,1.15);
\draw[prop] (D) -- ++(-1.10,-1.65);
\draw[prop] (D) -- ++( 1.10,-1.65);

\coordinate (XD) at ($(D)!0.52!($(D)+(-1.10,-1.65)$)$);
\node[ins] at (XD) {$\times$};
\node[lab] at ($(XD)+(0.22,-0.45)$) {$\mathcal F$};

\node[biglab] at (0.35,-4.25) {$+$};

\coordinate (E) at (2.2,-4.0);
\node[wzwvertex] at (E) {};
\draw[prop]  (E) -- ++(0,0.5);
\draw[prop]  (E) -- ++(-1.10,-1.65);
\draw[dprop] (E) -- ++( 1.10,-1.65);

\coordinate (XE) at ($(E)!0.52!($(E)+(1.10,-1.65)$)$);
\node[ins] at (XE) {$\times$};
\node[lab] at ($(XE)+(0.18,-0.52)$) {$\mathcal F$};

\node[biglab] at (4.35,-4.25) {$+$};

\coordinate (F) at (6.1,-4.0);
\node[wzwvertex] at (F) {};
\draw[prop]  (F) -- ++(0,0.5);
\draw[prop]  (F) -- ++(-1.10,-1.65);
\draw[dprop] (F) -- ++( 1.10,-1.65);

\coordinate (XF) at ($(F)!0.52!($(F)+(-1.10,-1.65)$)$);
\node[ins] at (XF) {$\times$};
\node[lab] at ($(XF)+(0.25,-0.45)$) {$\mathcal F$};

\node[lab] at (-1.8,-6.4) {$S_H^{(3)}\times S^{[2]}_{H,\mathcal{F}^{(2,0)}}\ \text{part}$};
\draw[brace,decoration={brace,mirror,amplitude=6pt}]
(1.0,-6.15) -- (7.3,-6.15)
node[midway,yshift=-0.55cm] {$\delta\rho^{[2]}\times S^{[2]}_{H,\mathcal{F}^{(2,0)}}\ \text{part}$};

\node[biglab] at (8.8,-4.25) {$+\cdots$};

\end{tikzpicture}
}}
\end{equation}
The Feynman rules and explicit momentum-space expressions for all diagrams are collected in
 Appendix~\ref{app: Feynman_R_and_D}. Diagrams containing dashed lines involve angular derivatives $\partial_\theta$, which modify the pole structure and make the corresponding contour integrations considerably more involved. For our purposes, however, evaluating all of these contributions is unnecessary. We instead isolate the contribution $\left\langle \delta\rho\,\delta\rho\,\delta\rho \right\rangle_{S^{[2]}_{H,\mathcal{F}^{(2,0)}}\times S^{[3]}_{H,\mathrm{free}}}$, which carries an explicit dependence on the dispersion curvature $\epsilon''(p_F)$ inherited from the free cubic Hamiltonian vertex. Recall that we take the Landau function to depend only on angles and set $\mathcal{F}^{(n,m)}=0$ for $n\geq3$ or $m>0$. Under these assumptions, this is the only contribution at this order that depends explicitly on $\epsilon''(p_F)$, and therefore it cannot be cancelled by the remaining diagrams. A nonvanishing contribution from this term is consequently sufficient to establish the existence of an additional nonanalytic momentum structure distinct from the noninteracting one. The details of the calculation are given in
Appendix~\ref{app: epsillon_dprine_derivation}. Here, we simply present the final expression for the contribution of the $\epsilon''(\mathbf{p})$ term to the equal-time three-point density correlation function:

\begin{align}
    \label{eqn: epsilon double prime_1}
     \int_{\omega_{k},\omega_{k_1}}\langle
\delta \rho_{\mathbf{k}}\,
\delta \rho_{\mathbf{k}_1}\,
\delta \rho_{\mathbf{k_2}}
\rangle_{\mathcal{O}(\mathcal{F})}^{\epsilon^{\prime\prime}}
&=\frac{ p_{F}\epsilon^{\prime\prime}(p_{F})}{12\pi v_{F}}\Bigg[-
|k_1|\,|k|
\int_{\theta,\theta'}\,
\mathcal{F}^{(2,0)}_{\theta,\theta'}
\frac{
k_{2n}k_{2n'}
(\hat{k}_{1}\cdot\hat{n}^{'})(\hat{k}\cdot\hat{n}^{'})
}{
\left[k_{2n}-k_{2n'}\right]^2
}
\notag\\
&\qquad\times
\bigg[
\Theta\!\left(-\mathrm{sgn}\!\left(k_{2n}\right)\right)
\Theta\!\left(-\mathrm{sgn}\!\left(k_{1n'}\right)\right)
\Theta\!\left(-\mathrm{sgn}\!\left(k_{n'}\right)\right)
+
\Theta\!\left(\mathrm{sgn}\!\left(k_{2n}\right)\right)
\Theta\!\left(\mathrm{sgn}\!\left(k_{1n'}\right)\right)
\Theta\!\left(\mathrm{sgn}\!\left(k_{n'}\right)\right)
\bigg]\notag\\
&\qquad-|k||k_2|\int_{\theta,\theta'} \,\frac{\mathcal{F}^{(2,0)}_{\theta,\theta'}(\hat{k}.\hat{n})(\hat{k_2}.\hat{n})k_{1n}k_{1n'}}{(k_{1n'}-k_{1n})^2}\notag\\&\qquad\quad\times\bigg[\Theta(-\text{sgn}(k_{2n}))\Theta(-\text{sgn}(k_{n}))\Theta(-\text{sgn}(k_{1n'}))+\Theta(\text{sgn}(k_{2n}))\Theta(\text{sgn}(k_{n}))\Theta(\text{sgn}(k_{1n'}))\bigg]\notag\\
&\qquad-|k_{1}||k_2|\int_{\theta,\theta'} \,\frac{\mathcal{F}^{(2,0)}_{\theta,\theta'}k_{n}k_{n'}(\hat{k_1}.\hat{n})(\hat{k_2}.\hat{n})}{(k_{n'}-k_{n})^2}\notag\\
&\qquad\quad\times\bigg[\Theta(-\text{sgn}(k_{2n}))\Theta(-\text{sgn}(k_{1n}))\Theta(-\text{sgn}(k_{n'}))+\Theta(\text{sgn}(k_{2n}))\Theta(\text{sgn}(k_{1n}))\Theta(\text{sgn}(k_{n'}))\bigg]\Bigg]\notag\\
&=
-\frac{p_F \epsilon''(p_{F})}{12\pi v_{F}}
|k|^2 \mathcal{F}_0\,\mathcal{I}(\varphi-\varphi_1).
\end{align}
To explicitly demonstrate that the integral in
Eq.~\eqref{eqn: epsilon double prime_1} is nonzero, we consider the simplest
case in which both the Landau interaction and the band-curvature factor are
angle independent, taking $\mathcal{F}^{(2,0)}_{\theta,\theta'}=\mathcal{F}_0$, and $|\mathbf{k}|=|\mathbf{k}_1|\equiv k$. 
For an isotropic dispersion, the band curvature enters only through its
value at the Fermi surface, $\epsilon''(p_F)$, which is independent of the
angular integration variables and has therefore been factored out in
Eq.~\eqref{eqn: epsilon double prime_1}. This should be distinguished from the Galilean case, for which
$\epsilon''(p_F)=v_F/p_F$, so that the ratio
$p_F\epsilon''(p_F)/v_F$ is fixed to unity and the band curvature is no
longer an independent coefficient of the effective Hamiltonian.
Our isolation argument then no longer applies, and a definite conclusion
would require evaluating the remaining
$\mathcal{O}(\mathcal{F}^{(2,0)})$ diagrams.

\begin{figure}[H]
    \centering
    \includegraphics[width=1.05\linewidth]{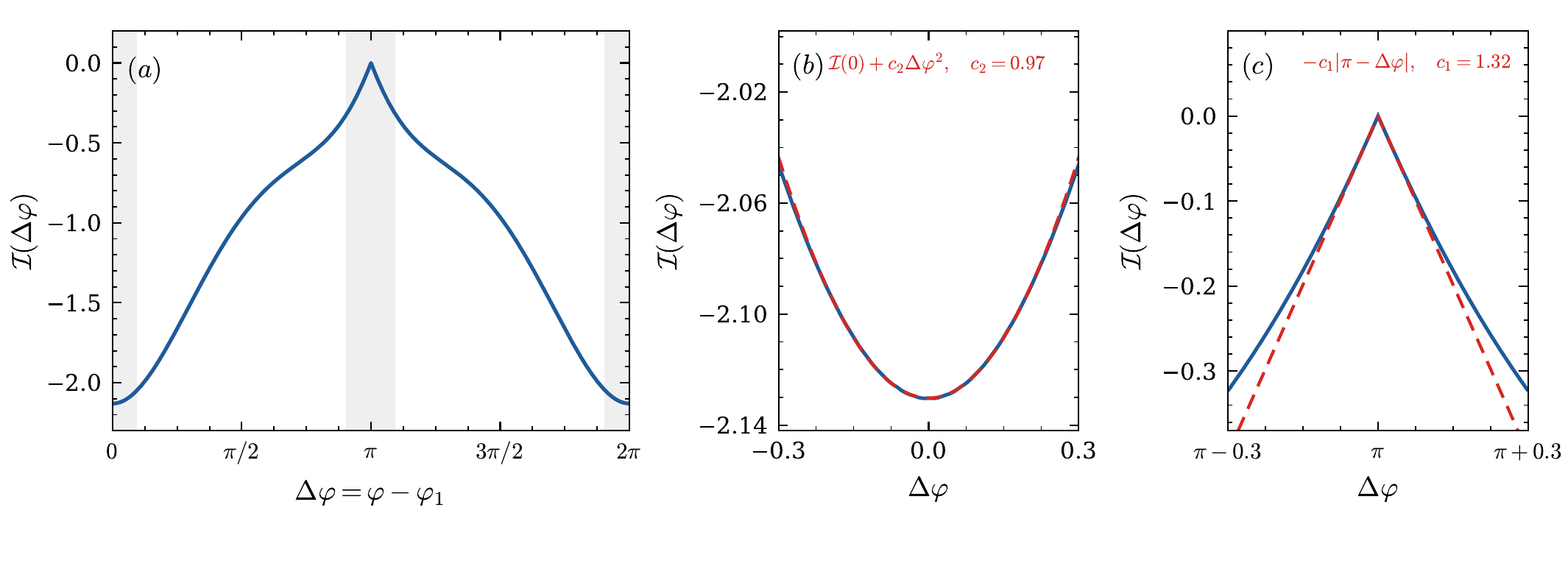}
    \captionsetup{font=small}
  \caption{The $\epsilon''(p_F)$ contribution to the equal-time three-point
density correlator, Eq.~\eqref{eqn: epsilon double prime_1}, as a function of the relative
orientation $\Delta\varphi=\varphi-\varphi_1$ of the external momenta, for
$|\vec{k}|=|\vec{k}_1|$ and angle-independent Landau interaction
$\mathcal{F}^{(2,0)}_{\theta\theta'}=\mathcal{F}_0$.
(a) Full angular range; shading marks the regions enlarged in (b) and (c).
(b) Near the parallel configuration, $\Delta\varphi=0$,
$\mathcal{I}(\Delta\varphi)=\mathcal{I}(0)+c_2\Delta\varphi^2$ with
$c_2\simeq1.0$ (dashed).
(c) Near the antiparallel configuration, $\Delta\varphi=\pi$,
$\mathcal{I}(\Delta\varphi)\approx-c_1|\pi-\Delta\varphi|$ with
$c_1\simeq1.32$ (dashed).}
\label{fig:epsilonpp_integral}
\end{figure}

With these assumptions the correlator reduces to the dimensionless integral
$\mathcal{I}(\varphi-\varphi_1)$, which we evaluate numerically in Figure \ref{fig:epsilonpp_integral} as a function of
the relative orientation of the external momenta. Denoting by $\varphi$ and
$\varphi_1$ the angles that $\mathbf{k}$ and $\mathbf{k}_1$ make with the
$x$-axis, rotational invariance implies that $\mathcal{I}$ depends only on the
difference $\varphi-\varphi_1$, so we may fix $\varphi_1=\pi/6$ without loss of
generality and vary $\varphi$. The resulting nonzero values of $\mathcal{I}$
show explicitly that the $\epsilon''(p_{F})$ contribution is finite.
As shown in Fig.~\ref{fig:epsilonpp_integral}, the integral is smooth near the
parallel configuration, $\Delta\varphi=0$, where it exhibits a quadratic
(parabolic) dependence,
\begin{align}
    \mathcal{I}(\Delta\varphi)
    \approx \mathcal{I}(0)+c_2\Delta\varphi^2,
    \qquad c_2\simeq 1.0.
\end{align}
This quadratic behavior can also be established analytically by expanding
the angular integrals near $\Delta\varphi=0$; the details are given in
Appendix~\ref{App: smooth}.

In contrast, at the antiparallel configuration, $\Delta\varphi=\pi$, it
vanishes linearly and develops a nonanalytic cusp,
\begin{align}
    \mathcal{I}(\Delta\varphi)
    \approx -c_1\left|\pi-\Delta\varphi\right|,
    \qquad c_1\simeq 1.32.
\end{align}
The vanishing at the antiparallel configuration is necessary for consistency
with the Ward identities. The linear cusp observed numerically can also be
established analytically by expanding the angular integrals near
$\Delta\varphi=\pi$; the details are given in
Appendix~\ref{app: cusp}. We investigate the corresponding symmetry
constraints in the next section.


\subsection{Symmetry Constraints on the Nonanalytic Structure}

We now examine how the nonanalytic contribution proportional to 
$\epsilon''(p_F)$ is consistent with permutation symmetry and charge 
conservation. Writing
\begin{equation}
    S_3(\mathbf{k},\mathbf{k}_1)
    =
    \widetilde{S}_3(\mathbf{k},\mathbf{k}_1,\mathbf{k}_2)\,
    \delta(\mathbf{k}+\mathbf{k}_1+\mathbf{k}_2),
\end{equation}
the reduced correlator $\widetilde{S}_3$ must be invariant under permutations 
of the three external momenta. Charge conservation further requires the Ward 
identity
\begin{equation}
    \widetilde{S}_3
    (\mathbf{k}=0,\mathbf{k}_1,\mathbf{k}_2=-\mathbf{k}_1)
    =0,
\end{equation}
together with its permutations. Before considering the explicit interaction correction, let us first ask 
whether an analytic contribution can occur at quadratic order in the external 
momenta. For a rotationally invariant and parity-invariant system, the most 
general permutation-symmetric analytic scalar at order $k^2$ can be written as
\begin{equation}
    \widetilde{S}_3^{\mathrm{analytic},(2)}
    =
    A\sum_{i=0}^{2}k_i^2
    +
    B\sum_{i<j}\mathbf{k}_i\cdot\mathbf{k}_j
\end{equation}
Here, for ease of notation $\mathbf{k}_{0}\equiv \mathbf{k}$. Momentum conservation implies
\begin{equation}
    \sum_{i=0}^{2}k_i^2
    +
    2\sum_{i<j}\mathbf{k}_i\cdot\mathbf{k}_j
    =0,
\end{equation}
so there is only one independent analytic structure at this order. We may 
therefore write
\begin{equation}
    \widetilde{S}_3^{\mathrm{analytic},(2)}
    =
    C\sum_{i=0}^{2}k_i^2 .
\end{equation}
Applying the Ward identity, for example by taking $\mathbf{k}_2=0$ and 
$\mathbf{k}=-\mathbf{k}_1$, gives
\begin{equation}
    \widetilde{S}_3^{\mathrm{analytic},(2)}
    =
    2Ck_1^2 .
\end{equation}
Charge conservation therefore requires $C=0$. Hence there exists no nonzero analytic, permutation-symmetric contribution to the density 
three-point function at order $k^2$. This observation is consistent with the nonzero free-fermion result,
\begin{equation}
S_3^{\mathrm{free}}(\mathbf{k},\mathbf{k}_1)
=
\frac{|\mathbf{k}\times\mathbf{k}_1|}{(2\pi)^2},
\end{equation}
which is nonanalytic because of the absolute value, despite scaling
quadratically with momentum. There is, however, an important distinction
between the nonanalyticity of the free result and that generated by the
band-curvature correction. To make this distinction transparent, consider
the equal-magnitude slice $|\mathbf{k}|=|\mathbf{k}_1|\equiv k$ and define
$\Delta\varphi\equiv\varphi-\varphi_1$. The free contribution becomes
\begin{equation}
S_3^{\mathrm{free}}
\propto
k^2|\sin\Delta\varphi|.
\end{equation}
It is therefore nonanalytic at both collinear configurations,
\begin{equation}
\Delta\varphi=0,
\qquad
\Delta\varphi=\pi,
\end{equation}
with local behavior $|\sin\Delta\varphi|\sim|\Delta\varphi|$ near the
parallel point and $|\sin\Delta\varphi|\sim|\Delta\varphi-\pi|$ near the
antiparallel point. In contrast, the $\epsilon''(p_F)$ contribution has a
different angular dependence: it remains finite and smooth at
$\Delta\varphi=0$, but vanishes with a cusp at $\Delta\varphi=\pi$.
Thus, although both contributions are nonanalytic, they exhibit distinct
angular singularity structures, and the $\epsilon''(p_F)$ term is not simply
proportional to the noninteracting result. Schematically, the $\epsilon''(p_F)$ contribution in
Eq.~\eqref{eqn: epsilon double prime_1} takes the form
\begin{align}
\widetilde{S}_3(\mathbf{k},\mathbf{k}_1,\mathbf{k}_2)
={}&
|\mathbf{k}||\mathbf{k}_1|\,f_{01}
+
|\mathbf{k}_1||\mathbf{k}_2|\,f_{12}
+
|\mathbf{k}||\mathbf{k}_2|\,f_{02}
+\mathcal{O}(k^3),
\end{align}
where each $f_{ij}$ depends only on the relative orientations
$\varphi_{ij}\equiv\varphi_i-\varphi_j$ of the external momenta and contains
the step-function constraints arising from the pole prescription. The
nonanalyticity is manifest in two places. First, the prefactors
$|\mathbf{k}_i|$ are not analytic in the momentum components. Second, the step
functions make the angular domain of integration depend on the external
directions through sign conditions, so the $f_{ij}$ are themselves non-smooth
where that domain collapses. 

As a consistency check, we now verify directly that this nonanalytic
$\epsilon''(p_{F})$ contribution satisfies the Ward identity. Consider the
limit $\mathbf{k}_2\rightarrow0$. The second and third terms vanish
explicitly because they are proportional to $|\mathbf{k}_2|$. The first term,
however, contains no explicit factor of $|\mathbf{k}_2|$, so its vanishing
must instead follow from the pole constraints. Momentum conservation in this
limit gives $\mathbf{k}_1=-\mathbf{k}$. The factors
$\Theta\left[-\mathrm{sgn}(k_{n'})\right]
\Theta\left[-\mathrm{sgn}(k_{1n'})\right]$
appearing in the first term require $k_{n'}$ and $k_{1n'}$ to have the same
sign. For antiparallel external momenta, however,
$k_{1n'}=-k_{n'}$, so these conditions cannot be simultaneously satisfied; therefore, the contribution vanishes. By permutation symmetry, the same conclusion follows when either of the other
external momenta is taken to zero. The $\epsilon''(p_F)$ contribution is therefore fully consistent with charge
conservation. More generally, the Ward identity and permutation symmetry rule
out any nonzero analytic contribution at order $k^2$, but do not uniquely
select the noninteracting topological form. Distinct nonanalytic contributions,
controlled by independent microscopic parameters, can occur at the same order
in momentum.
\section{Discussion}
\label{Sec 4}
In this work, we study how interactions modify topology-probing equal-time
density correlations in a two-dimensional Fermi liquid using the
coadjoint-orbit formulation. This framework provides a systematic organization
of generalized Landau interactions $\mathcal{F}^{(n,m)}$, while mapping the
fermionic loop expansion onto simpler tree-level diagrams in the bosonized
variables. The formalism can also be extended to include internal degrees of
freedom, such as spin~\cite{Mehta2023}. The main technical difficulty instead lies in the frequency contour integrals,
owing to the subtle pole structure and the associated singular contributions. 

We first verify that the coadjoint-orbit formulation reproduces the
noninteracting result of Ref.~\cite{Kane} for a simply connected Fermi sea.
At first order in the generalized Landau interaction
$\mathcal{F}^{(2,0)}$, we isolate a contribution proportional to the band
curvature $\epsilon''(p_F)$. For a general isotropic dispersion,
$v_F=\epsilon'(p_F)$ and $\epsilon''(p_F)$ are independent coefficients of
the effective Hamiltonian. Under these assumptions, no other contribution at this order carries an explicit dependence on $\epsilon''(p_F)$, so the term can be isolated from the remaining diagrams.
\footnote{For a Galilean dispersion,
$\epsilon''(p_F)=v_F/p_F$, so the band curvature is not an independent
coefficient. In that case, establishing the full correction requires
evaluating the remaining
$\mathcal{O}(\mathcal{F}^{(2,0)})$ diagrams.}

The band-curvature contribution is nonanalytic, but with an angular structure
qualitatively different from that of the noninteracting term: it remains finite and smooth at the parallel
configuration and vanishes with a cusp at the antiparallel one. This should be distinguished from the singularity analyzed in Refs.~\cite{Tam2026Singular, Kane2026ThreePoint}. There the object of
interest is the discontinuity in the slope of $S_3$ as a function of the angle
between the external momenta, which is sharply defined in the long-wavelength
collinear limit, and whose coefficient is renormalized by Landau parameters. A
contribution that is smooth at the parallel configuration does not affect that
slope discontinuity; it shifts the value of the correlator without altering the
kink. The two statements therefore refer to different features of $S_3$, and the
band-curvature term identified here is a structure that the collinear limit
projects out. The systems also differ: Ref.~\cite{Kane2026ThreePoint} computes the
total-density correlator of a spin-$1/2$ gas with a contact interaction acting
only between opposite spins, for which there is no first-order counterpart of
the single-component Landau interaction considered here. Our result shows that, away from the collinear limit and for a general isotropic dispersion, the
equal-time correlator contains a second nonanalytic structure whose size is set
by the dimensionless ratio $p_F\epsilon''(p_F)/v_F$, so that the angular profile
of $S_3$ is not fixed by the Landau parameters alone.

Our result does not contradict the close agreement between the measured normalized correlator $S_3$
and the ideal-gas prediction reported in Ref.~\cite{Daix2025_FermiSea}. That experiment extracts $S_3$ from configurations near $\Delta \varphi=\pi/2$,
where the correction identified here is small relative to the
noninteracting term, which is largest there. By contrast, the relative
importance of the correction is enhanced closer to the collinear regime
$\Delta\varphi\to0$, which was not the primary regime explored in the
experiment. The measured correlator is also that
of a single spin component of a two-component gas whose interactions act only
between opposite spins, rather than the total density to which the
single-component Landau interaction considered here applies. Finally, that experiment realizes a Galilean-invariant gas, for which
$p_F\epsilon''(p_F)/v_F$ is fixed to unity and cannot be tuned; for a general
dispersion it is a free parameter. The effect should therefore be sought in the
total-density correlator of a two-component gas, or in lattice systems, where
$\epsilon''(p_F)$ can be varied through filling. Reaching the near-collinear regime additionally requires access to smaller momenta than in Ref.~\cite{Daix2025_FermiSea}, since the validity condition for the topological formula derived there becomes increasingly restrictive as the momenta approach collinearity, and is recovered only by reducing their magnitudes. The accessible momentum resolution is set by the real-space cutoff of the correlation measurement, so larger clouds would open this window directly.

Our result shows that interactions do more than renormalize the nonanalytic structure already present in the noninteracting theory. We find that, already at $\mathcal{O}(\mathcal{F})$, the density three-point function contains an additional nonanalytic contribution proportional to the independent band-curvature parameter $\epsilon''(p_F)$, with momentum dependence distinct from that of the noninteracting result. The existence of such an independent contribution is sufficient to show that the full interacting density three-point function is not determined by Fermi-sea topology alone. More generally, higher orders in the effective theory introduce additional dispersion coefficients and generalized Landau parameters, thereby allowing further nonanalytic structures consistent with the symmetry constraints. Thus, although the topology-sensitive nonanalyticity of the noninteracting theory can persist in the presence of interactions, it no longer uniquely characterizes the full density three-point
function, which acquires additional interaction-dependent structures not
fixed by Fermi-sea topology.

The search for interaction-robust probes of Fermi-surface topology therefore
remains open, and the one-dimensional case suggests where to look. There, the
quantities that survive interactions are those fixed by the symmetry algebra of
the low-energy theory rather than by the Hamiltonian. The density two-point
function is renormalized by the Luttinger parameter, while the stress-tensor
correlator is fixed by the central charge, and the Euler characteristic
remains visible and free of interaction dependence. This points to a broader
possibility. Interaction-independent topological information in higher
dimensions may reside in observables controlled by the kinematic sector of the
theory, such as the Wess-Zumino-Witten term, rather than in density correlations
alone. The coadjoint-orbit formulation is well suited to that search.

\section{Acknowledgments}
We thank Pok Man Tam, Luca V. Delacretaz, Jack Farrell, Xiaoyang Huang and Victor Gurarie for helpful discussions and valuable comments on the manuscript. Large language models, including Claude Fable 5, Claude Opus 5, and GPT-5.6, were used as auxiliary tools for brainstorming, language and grammar refinement, and manuscript auditing. A.P. and A.L. acknowledge support from the National Science Foundation under Grant No. DMR-2145544. U.M. acknowledges support from a Simons Postdoctoral Fellowship in Ultra-Quantum Matter and from the Gordon and Betty Moore Foundation under Grant No. GBMF10279.
\appendix


\section{Density Three-Point Function in a Two-Dimensional Fermi Liquid}

\subsection{Feynman Rules and Diagrammatic Conventions}
\label{app: Feynman_R_and_D}
In this appendix, we collect the relevant Feynman rules and summarize the diagrams required to evaluate the density three-point correlation function within the coadjoint-orbit effective theory introduced in the main text. The Gaussian part of the action, $S^{[2]}_{G}$, determines the free propagator of the theory in momentum space:
\begin{equation}
\label{eqn: Two point correlation}
\begin{tikzpicture}[baseline=(current bounding box.center)]
    \draw[thick] (0,0) -- (1.8,0);
\end{tikzpicture}
\quad\equiv\quad
\langle\phi^{\theta}_{\mathbf{k}}\phi^{\theta'}_{\mathbf{k_1}}\rangle_{0}
=
\frac{i}{p_{F}}
\frac{\delta_{\theta,\theta'}}
{k_{n}(\omega_{\mathbf{k}}-v_{F}k_{n})}
(2\pi)^2\delta_{\mathbf{k+k_1}}
(2\pi)\delta_{\omega_{\mathbf{k}}+\omega_{\mathbf{k_1}}}
\end{equation}
where $k_{n}=\mathbf{k.n_{\theta}}$.
Throughout this work, we adopt the Feynman pole prescription. Thus, poles at $\omega=v_F k_n$ are understood as $\omega=v_F k_n-i\epsilon\,\mathrm{sgn}(\omega)$. To avoid clutter, this prescription will be kept implicit in all subsequent expressions. The full propagator of the theory can be written as an infinite series containing contributions from all the non-Gaussian terms of the full action. Here we note only that the leading $\mathcal{O}(\mathcal{F})$ correction to the two-point function comes from the interacting part of the Gaussian action 
$S^{\mathcal{O}(\mathcal{F})}_{H^{(2)}}$. Diagrammatically, this contribution is represented by a cross insertion on the propagator line.
\begin{equation}
\label{eqn: full propagator}
\begin{aligned}
\begin{tikzpicture}[baseline=(current bounding box.center), line cap=round, line join=round, scale=1.0]
\tikzset{
    prop/.style={line width=1.2pt},
    dprop/.style={double, line width=1.2pt, double distance=1.2pt},
    cross/.style={circle, draw, inner sep=1.8pt, line width=0.8pt},
    lab/.style={font=\small}
}
    \draw[dprop] (0,0) -- (1.8,0);
\end{tikzpicture}
\quad\equiv\quad
\langle \phi^{\theta}_{\mathbf{k}}\phi^{\theta'}_{\mathbf{k}_1}\rangle
&=
\begin{tikzpicture}[baseline=(current bounding box.center), line cap=round, line join=round, scale=1.0]
\tikzset{
    prop/.style={line width=1.2pt},
    cross/.style={circle, draw, inner sep=1.8pt, line width=0.8pt},
    lab/.style={font=\small}
}
    \draw[prop] (0,0) -- (1.7,0);
    \node[lab] at (0.85,0.75) {$\langle \phi\phi\rangle_{0}$};

    \node[lab] at (2.15,0) {$+$};

    \draw[prop] (2.8,0) -- (5.4,0);
    \node[cross] at (4.1,0) {$\times$};
    \node[lab] at (4.1,-0.55) {$\mathcal{F}$};
    \node[lab] at (4.1,0.9) {$\langle \phi\phi\rangle_{1}$};

    \node[lab] at (5.85,0) {$+$};
    \node[lab] at (6.35,0) {$\cdots$};
\end{tikzpicture}
\\[1em]
&=
\frac{i}{p_{F}}\frac{\delta_{\theta,\theta'}}{k_{n}(\omega-v_{F}k_{n})}(2\pi)^2\delta_{\mathbf{k+k_1}}(2\pi)\delta_{\omega_{\mathbf{k}}+\omega_{\mathbf{k_1}}}
\nonumber\\
&\quad+
\frac{i v_{F}}{p_{F}}\frac{\mathcal{F}_{\theta,\theta'}}{(\omega-v_{F}k_{n})(\omega-v_{F}k_{n'})}(2\pi)^2\delta_{\mathbf{k+k_1}}(2\pi)\delta_{\omega_{\mathbf{k}}+\omega_{\mathbf{k_1}}}
+\mathcal{O}(\mathcal{F}^2)+\cdots .
\end{aligned}
\end{equation}
The number density in this formalism is given by
\begin{align}
\rho(t,\mathbf{x})
= \int_{\mathbf{p}} f_{\phi}(t,\mathbf{x},\mathbf{p})
\quad \Rightarrow \quad
\delta \rho(t,\mathbf{x})
= \frac{p_F}{(2\pi)^2}\int_{\theta}
\left[
\nabla_n \phi
+ \frac{1}{2p_F}\nabla_s\!\left(\partial_{\theta}\phi\,\nabla_n\phi\right)
+ \mathcal{O}(\phi^3)
\right].
\end{align}
In the above, corrections to $\delta\rho$ beyond quadratic order are suppressed by powers of $1/p_F$. Since the relevant low-temperature quasiparticle excitations correspond to small deformations of the Fermi surface, 
$\delta k \ll p_F$, these curvature-suppressed contributions are subleading and will be neglected \cite{Umang}. 
With this approximation, the three-point density-density correlation function in momentum space is given by
\begin{align}
\label{eqn: three-point density correlation expansion}
    \langle \delta \rho(k,\omega)\delta\rho(k_1,\omega_1)\delta\rho(k_2,\omega_2)\rangle &= i\frac{p^3_{F}}{(2\pi)^6}\int_{\theta,\theta',\theta''}k_{n}k_{1n'}k_{2n''}\langle\phi^{\theta}_{\mathbf{k}}\phi^{\theta'}_{\mathbf{k_1}}\phi^{\theta''}_{\mathbf{k_2}}\rangle+\notag\\
    &\qquad+ \Big(\frac{p^2_{F}}{2(2\pi)^6}\int_{\theta,\theta',\theta'',p,\omega_{p}}k_{n}k_{1n'}k_{2s''}\big(k_2-p\big)_{n''}\langle\phi^{\theta}_{\mathbf{k}}\phi^{\theta'}_{\mathbf{k_1}}\partial_{\theta''}\phi^{\theta''}_{\mathbf{p}}\phi^{\theta''}_{\mathbf{k_2-p}}\rangle+\text{5-permutations}\Big)+\cdots
\end{align}
At the level of field counting alone, one might expect the first term in Eq.~\eqref{eqn: three-point density correlation expansion}, involving the three-point function of $\phi$, to dominate over the second term, which involves a four-point function of $\phi$. However, the relevant $1/p_F$ power counting shows that this expectation is not correct. The leading contribution to $\langle \phi^{\theta}_{\mathbf{k}}\phi^{\theta'}_{\mathbf{k}_1}\phi^{\theta''}_{\mathbf{k}_2}\rangle$ arises from the cubic vertices in the action, namely $\langle \phi^{\theta}_{\mathbf{k}}\phi^{\theta'}_{\mathbf{k}_1}\phi^{\theta''}_{\mathbf{k}_2}S^{[3]}_{\mathrm{WZW}}\rangle_0$ or $\langle \phi^{\theta}_{\mathbf{k}}\phi^{\theta'}_{\mathbf{k}_1}\phi^{\theta''}_{\mathbf{k}_2}S^{[3]}_{H}\rangle_0$, and is therefore given by a product of three free propagators. Since each free propagator carries a factor of $1/p_F$, this contribution scales as $p_F^3(1/p_F)^3 \sim \mathcal{O}(1)$. On the other hand, the four-point function appearing in the second term of Eq.~\eqref{eqn: three-point density correlation expansion} already has a nonzero Wick contraction at Gaussian order, for example $\langle \phi^{\theta}_{\mathbf{k}}\phi^{\theta'}_{\mathbf{k}_1}\partial_{\theta''}\phi^{\theta''}_{\mathbf{p}}\phi^{\theta''}_{\mathbf{k}_2-\mathbf{p}}\rangle_0 \sim\allowbreak \langle \phi^{\theta}_{\mathbf{k}}\partial_{\theta''}\phi^{\theta''}_{\mathbf{p}}\rangle_0 \langle \phi^{\theta'}_{\mathbf{k}_1}\phi^{\theta''}_{\mathbf{k}_2-\mathbf{p}}\rangle_0 + \text{permutations}$. This term is a product of two free propagators and hence scales as $p_F^2(1/p_F)^2 \sim \mathcal{O}(1)$. Thus, despite involving a higher-point correlator of $\phi$, the second term in Eq.~\eqref{eqn: three-point density correlation expansion} contributes at the same order in the $1/p_F$ expansion as the first term. Consequently, both terms must be retained in the leading evaluation of the density three-point function.
We now evaluate these contributions one by one using diagrammatic perturbation theory.
\begin{align}
    \langle \phi^{\theta}_{\mathbf{k}}\phi^{\theta'}_{\mathbf{k_1}}\phi^{\theta''}_{\mathbf{k_2}}\rangle&= \langle \phi^{\theta}_{\mathbf{k}}\phi^{\theta'}_{\mathbf{k_1}}\phi^{\theta''}_{\mathbf{k_2}}\: \exp\big[-i(S^{[2]}_{H,\mathcal{F}^{(2,0)}}+S^{[3]}_{H}+S^{[3]}_{WZW})\big]\rangle^{\text{connected}}_{0}\notag\\
    &=-i\langle \phi^{\theta}_{\mathbf{k}}\phi^{\theta'}_{\mathbf{k_1}}\phi^{\theta''}_{\mathbf{k_2}} S^{[3]}_{WZW}\rangle_{0} -i\langle \phi^{\theta}_{\mathbf{k}}\phi^{\theta'}_{\mathbf{k_1}}\phi^{\theta''}_{\mathbf{k_2}} S^{[3]}_{H}\rangle_{0}-\langle \phi^{\theta}_{\mathbf{k}}\phi^{\theta'}_{\mathbf{k_1}}\phi^{\theta''}_{\mathbf{k_2}}S^{[2]}_{H,\mathcal{F}^{(2,0)}}S^{[3]}_{WZW,\text{free}}\rangle-\langle \phi^{\theta}_{\mathbf{k}}\phi^{\theta'}_{\mathbf{k_1}}\phi^{\theta''}_{\mathbf{k_2}}S^{[2]}_{H,\mathcal{F}^{(2,0)}}S^{(3)}_{H,\text{free}}\rangle  +\cdots
\end{align}
\[
\begin{tikzpicture}[
    scale=0.9,
    lvertex/.style={circle, fill=black, inner sep=1.2pt},
    kvertex/.style={rectangle, fill=red, draw=red, inner sep=1.8pt},
    ins/.style={circle,draw,inner sep=0.05pt},
    lab/.style={font=\small}
]

\draw[ultra thick] (-1.2,0) -- (-1.2,1.1);
\draw[ultra thick] (-1.2,0) -- (-1.8,-1.1);
\draw[ultra thick] (-1.2,0) -- (-0.4,-1);

\node at (0.2,-0.3) {$=$};

\coordinate (A) at (2,0);
\node[lvertex] at (A) {};
\draw (A) -- (2,1.2);
\draw (A) -- (1.1,-1.0);
\draw[dashed] (A) -- (3.0,-0.8);

\node[lab] at (2.5,0.65) {$\langle \phi\phi\rangle_0$};
\node[lab] at (1.06,-0.3) {$\langle \phi\phi\rangle_0$};
\node[lab] at (3.1,-0.25) {$\langle \phi\partial_\theta\phi\rangle_0$};

\draw[decorate,decoration={brace,amplitude=5pt}]
  (1.0,1.25) -- (3.15,1.25);
\node at (2.1,2.05) {$O(1)$};

\draw[decorate,decoration={brace,amplitude=5pt,mirror}]
  (1.0,-1.45) -- (3.15,-1.45);
\node at (2.1,-1.85) {$S^{[3]}_{\mathrm{WZW}}\ \text{part}$};

\node at (4.0,-0.2) {$+$};

\coordinate (B) at (5.4,0);
\node[kvertex] at (B) {};
\draw (B) -- (5.4,1.2);
\draw (B) -- (4.6,-1.0);
\draw (B) -- (6.2,-1.0);

\node[lab] at (5.9,0.65) {$\langle\phi\phi\rangle_0$};
\node[lab] at (4.57,-0.35) {$\langle\phi\phi\rangle_0$};
\node[lab] at (6.27,-0.35) {$\langle\phi\phi\rangle_0$};

\draw[decorate,decoration={brace,amplitude=5pt}]
  (4.6,1.65) -- (6.2,1.65);
\node at (5.4,2.05) {$O(1)$};

\node at (7.0,-0.2) {$+$};

\coordinate (C) at (8.5,0);
\node[kvertex] at (C) {};
\draw (C) -- (8.5,1.2);
\draw (C) -- (7.8,-1.0);
\draw[dashed] (C) -- (9.5,-0.8);
\node[ins] at (8.5,0) {$\times$};
\node[lab] at (8.05,0) {$\mathcal{F}$};

\node[lab] at (8.9,0.65) {$\langle\phi\phi\rangle_0$};
\node[lab] at (7.7,-0.4) {$\langle\phi\phi\rangle_0$};
\node[lab] at (9.8,-0.25) {$\langle\phi\partial_\theta\phi\rangle_0$};

\draw[decorate,decoration={brace,amplitude=3pt}]
  (7.7,1.4) -- (9.5,1.4);
\node at (8.6,2.05) {$O(\mathcal{F}^1)$};

\draw[decorate,decoration={brace,amplitude=6pt,mirror}]
  (4.3,-1.6) -- (9.8,-1.6);
\node at (7.0,-2.05) {$S_H^{[3]}\ \text{part}$};

\node at (-0.1,-3.7) {$+$};

\coordinate (D) at (1.6,-3.5);
\node[lvertex] at (D) {};
\draw (D) -- (1.6,-2.3);
\draw (D) -- (0.8,-4.5);
\draw[dashed] (D) -- (2.8,-4.2);
\node[ins] at (1.1,-4.15) {$\times$};
\node[lab] at (1.4,-4.28) {$\mathcal{F}$};
\node[lab] at (2,-2.85) {$\langle\phi\phi\rangle_0$};
\node[lab] at (0.7,-3.8) {$\langle\phi\phi\rangle_1$};
\node[lab] at (2.85,-3.75) {$\langle\phi\partial_\theta\phi\rangle_0$};

\node at (3.6,-3.7) {$+$};

\coordinate (E) at (5.0,-3.5);
\node[lvertex] at (E) {};
\draw (E) -- (5.0,-2.3);
\draw (E) -- (4.2,-4.5);
\draw[dashed] (E) -- (6.2,-4.2);

\node[ins] at (5.65,-3.85) {$\times$};
\node[lab] at (5.2,-3.95) {$\mathcal{F}$};

\node[lab] at (5.5,-2.85) {$\langle\phi\phi\rangle_0$};
\node[lab] at (4.1,-4.0) {$\langle\phi\phi\rangle_0$};
\node[lab] at (6.3,-3.55) {$\langle\phi\partial_\theta\phi\rangle_1$};

\node at (7.8,-3.7) {$+$};

\draw[decorate,decoration={brace,amplitude=2pt,mirror}]
  (0.55,-5.0) -- (6,-5.0);
\node at (3.5,-5.5) {$S_{H,\mathcal{F}^{(2,0)}}^{[2]}S^{[3]}_{\mathrm{WZW},\text{free}}$ part $\mathcal{O}(\mathcal{F}^{1})$};

\coordinate (H) at (9.8,-3.5);
\node[kvertex] at (H) {};
\draw (H) -- (9.8,-2.3);
\draw (H) -- (9.0,-4.5);
\draw (H) -- (10.9,-4.3);

\node[ins] at (9.25,-4.15) {$\times$};
\node[lab] at (9.6,-4.28) {$\mathcal{F}$};

\node[lab] at (10.2,-2.95) {$\langle\phi\phi\rangle_0$};
\node[lab] at (9,-3.75) {$\langle\phi\phi\rangle_1$};
\node[lab] at (10.8,-3.85) {$\langle\phi\phi\rangle_0$};

\draw[decorate,decoration={brace,amplitude=6pt,mirror}]
  (8.75,-4.95) -- (11.0,-4.95);
\node at (9.8,-5.4) {$S_{H,\mathcal{F}^{(2,0)}}^{[2]}S^{[3]}_{\mathrm{H},\text{free}}$\:\text{part}\: $\mathcal{O}(\mathcal{F}^1)$};

\end{tikzpicture}
\]
Consider first the WZW diagram $S^{[3]}_{\mathrm{WZW}}$, which generates a three-point vertex since it is cubic in the field $\phi$. We denote the corresponding WZW cubic vertex by a round black dot 
(\(\tikz[baseline=-0.5ex]\node[circle,fill=black,inner sep=1.3pt]{};\)). Schematically,
$\langle
\phi^{\theta}_{\mathbf{k}}
\phi^{\theta'}_{\mathbf{k}_1}
\phi^{\theta''}_{\mathbf{k}_2}
S^{[3]}_{\mathrm{WZW}}
\rangle_{0}
\sim\allowbreak
\int_{\theta^*,\mathbf{p},\mathbf{p}_1}
\langle
\phi^{\theta}_{\mathbf{k}}
\phi^{\theta'}_{\mathbf{k}_1}
\phi^{\theta''}_{\mathbf{k}_2}
\phi^{\theta^*}_{\mathbf{p}}
\phi^{\theta^*}_{\mathbf{p}_1}
\partial_{\theta^*}
\phi^{\theta^*}_{-\mathbf{p}-\mathbf{p}_1}
\rangle_{0}
\sim\allowbreak
\int_{\theta^*,\mathbf{p},\mathbf{p}_1}
\langle
\phi^{\theta}_{\mathbf{k}}
\phi^{\theta^*}_{\mathbf{p}}
\rangle_{0}
\langle
\phi^{\theta'}_{\mathbf{k}_1}
\phi^{\theta^*}_{\mathbf{p}_1}
\rangle_{0}
\langle
\phi^{\theta''}_{\mathbf{k}_2}
\partial_{\theta^*}
\phi^{\theta^*}_{-\mathbf{p}-\mathbf{p}_1}
\rangle_{0}$.
Thus, the solid lines denote the usual propagators $\langle\phi\phi\rangle_{0}$, while the dashed line denotes the mixed contraction $\langle\phi\,\partial_{\theta}\phi\rangle_{0}$. We now turn to the contribution from $S^{[3]}_{H}$. This term also defines a three-point vertex, since it is cubic in $\phi$. We denote the corresponding three-point vertex by a red square (\(\tikz[baseline=-0.5ex]\node[rectangle,fill=red,draw=red,inner sep=1.8pt]{};\)). The second diagram comes from the noninteracting part of $S^{[3]}_{H}$, and therefore all three internal contractions are ordinary free propagators $\langle\phi\phi\rangle_{0}$. Hence, the corresponding diagram contains only solid lines. The third diagram comes from the interaction-dependent part, $S^{[3]}_{H,\mathcal{F}^{(2,0)}}$. In this case, the Landau parameter $\mathcal{F}$ appears explicitly in the cubic vertex. We therefore mark this vertex with a cross to indicate that this contribution is already of order $\mathcal{O}(\mathcal{F})$. For the class of diagrams contributing to
$\langle
\phi^{\theta}_{\mathbf{k}}
\phi^{\theta'}_{\mathbf{k}_1}
\phi^{\theta''}_{\mathbf{k}_2}
S^{[2]}_{H,\mathcal{F}^{(2,0)}}S^{[3]}_{\mathrm{WZW,\text{free}}}
\rangle$,
the cross associated with the Landau parameter $\mathcal{F}$ can appear either on a solid line, corresponding to the corrected propagator $\langle \phi\phi\rangle_{1}$, or on the dashed line, corresponding to the corrected mixed contraction $\langle \phi\,\partial_{\theta}\phi\rangle_{1}$. Similarly, for the class of diagrams arising from
$\langle
\phi^{\theta}_{\mathbf{k}}
\phi^{\theta'}_{\mathbf{k}_1}
\phi^{\theta''}_{\mathbf{k}_2}
S^{[2]}_{H,\mathcal{F}^{(2,0)}}S^{[3]}_{\mathrm{H},\text{free}}
\rangle$,
the Landau-parameter insertion can dress any one of the three ordinary propagators. Diagrammatically, this corresponds to placing the cross on any one of the three solid lines.
\vspace{1 em}
\par
\noindent
We now write the explicit form of all the above diagrams in momentum space:
\vspace{0.8em}

\begin{center}
\begin{tikzpicture}[line cap=round, line join=round, scale=1.0]

\tikzset{
    prop/.style={line width=1.2pt},
    dottedprop/.style={line width=1.1pt, dashed, dash pattern=on 4pt off 4pt},
    vertex/.style={circle, fill=black, inner sep=1.6pt},
    lab/.style={font=\small}
}

\coordinate (V) at (0,0);
\node[vertex] at (V) {};

\draw[prop] (V) -- (0,1.7);
\draw[prop] (V) -- (-1.25,-1.25);
\draw[dottedprop] (V) -- (1.35,-1.0);

\node[lab] at (0.55,1.00) {$\langle \phi\phi\rangle_{0}$};
\node[lab] at (-1.75,-0.35) {$\langle \phi\phi\rangle_{0}$};
\node[lab] at (1.95,-0.35) {$\langle \phi\,\partial_{\theta}\phi\rangle_{0}$};

\node[anchor=west, align=left] at (3.2,0.15) {%
\begin{minipage}{0.72\textwidth}
\begin{equation}
\label{eqn: WZW free diagram}
\begin{aligned}
=\, &\frac{- i}{3!}\frac{\delta_{\theta',\theta}}{p_F^3}
\frac{k_{1s}}
{\omega_k-v_Fk_n}
\frac{\big(\omega_k+2\omega_{k_1}\big)}
{k_{1n'}\big(\omega_{k_1}-v_Fk_{1n'}\big)}
\\[0.4em]
&\times
\partial_{\theta}\!\left(
\frac{\delta_{\theta''\theta}}
{k_{2n''}\big(\omega_{k_2}-v_Fk_{2n''}\big)}
\right)
(2\pi)^3
\delta_{k_2+k_1+k}\,
\delta_{\omega_{k_2}+\omega_{k_1}+\omega_k}
\\[0.4em]
&\qquad +(5\text{-permutations})
\end{aligned}
\end{equation}
\end{minipage}
};

\end{tikzpicture}
\end{center}
\vspace{0.8em}
\begin{center}
\begin{tikzpicture}[line cap=round, line join=round, scale=1.0]

\tikzset{
    prop/.style={line width=1.2pt},
    kvertex/.style={rectangle, fill=red, draw=red, inner sep=2.2pt},
    lab/.style={font=\small},
    smalllab/.style={font=\small}
}

\coordinate (V) at (0,0);
\node[kvertex] at (V) {};

\draw[prop] (V) -- (0,1.7);
\draw[prop] (V) -- (-1.25,-1.25);
\draw[prop] (V) -- (1.35,-1.0);

\node[smalllab] at (0.80,1.00) {$\langle \phi\phi\rangle_{0}$};
\node[smalllab] at (-1.95,-0.35) {$\langle \phi\phi\rangle_{0}$};
\node[smalllab] at (2.05,-0.35) {$\langle \phi\phi\rangle_{0}$};

\node[anchor=west, align=left] at (3.2,0.10) {%
\begin{minipage}{0.68\textwidth}
\begin{equation}
\label{eqn: H3_free}
\begin{aligned}
=\, &\Bigg[
\frac{i}{p_F^3}
\left(
\frac{v_F}{2}+p_F\epsilon''
\right)
\frac{k_{1n}(k+k_1)_n}{\omega_k-v_Fk_n}
\frac{\delta_{\theta,\theta'}}
     {k_{1n'}(\omega_{k_1}-v_Fk_{1n'})}
\frac{\delta_{\theta'',\theta}}
     {k_{2n''}(\omega_{k_2}-v_Fk_{2n''})}
\\[0.5em]
&\qquad\times
(2\pi)^3\delta_{k+k_1+k_2}\,
\delta_{\omega_k+\omega_{k_1}+\omega_{k_2}}
+(5\text{-permutations})
\Bigg]
\end{aligned}
\end{equation}
\end{minipage}
};

\end{tikzpicture}
\end{center}
\vspace{0.8em}

\begin{center}
\begin{tikzpicture}[line cap=round, line join=round, scale=1.0]

\tikzset{
    prop/.style={line width=1.2pt},
    dottedprop/.style={line width=1.1pt, dashed, dash pattern=on 4pt off 4pt},
    kvertex/.style={rectangle, fill=red, draw=red, inner sep=2.2pt},
    lab/.style={font=\small}
}

\coordinate (V) at (0,0);
\node[kvertex] at (V) {};

\draw[prop] (V) -- (0,1.7);
\draw[prop] (V) -- (-1.25,-1.25);
\draw[dottedprop] (V) -- (1.35,-1.0);

\node[draw, circle, inner sep=1.5pt] at (0.05,0.02) {$\times$};
\node[lab] at (0.62,0.02) {$\mathcal{F}$};

\node[lab] at (0.55,1.00) {$\langle \phi\phi\rangle_{0}$};
\node[lab] at (-1.75,-0.35) {$\langle \phi\phi\rangle_{0}$};
\node[lab] at (1.95,-0.35)
{$\langle \phi\,\partial_{\theta}\phi\rangle_{0}$};

\node[anchor=west, align=left] at (3.2,0.15) {%
\begin{minipage}{0.62\textwidth}

\begin{equation}
\label{eqn: WZW_Int_vertex}
\begin{aligned}
=&\Bigg(
\Big[
-\frac{i}{2}\frac{v_{F}}{p_{F}^{3}}
\mathcal{F}_{\theta,\theta''}
\frac{k_{s}}
{(\omega_{k}-v_{F}k_{n})
(\omega_{k_2}-v_{F}k_{2n''})}
\\
&\qquad\qquad\qquad\qquad\qquad\times
\partial_{\theta}\Big(
\frac{(2\pi)^3\delta_{\theta,\theta'}}
{k_{1n'}(\omega_{k_1}-v_{F}k_{1n'})}
\Big)
\delta_{k+k_1+k_2}
\delta_{\omega_k+\omega_{k_1}+\omega_{k_2}}
+(\theta\leftrightarrow\theta'')
\\[0.4em]
&\qquad
+\frac{i}{2}\frac{v_{F}}{p_{F}^{3}}
\mathcal{F}_{\theta',\theta''}
\frac{k_{n'}k_{1s'}}
{k_{1n'}(\omega_{k_1}-v_{F}k_{1n'})}
\frac{1}{\omega_{k_2}-v_{F}k_{2n''}}
\\
&\qquad\qquad\qquad\qquad\qquad\times
\partial_{\theta'}\Big(
\frac{(2\pi)^3\delta_{\theta,\theta'}}
{k_{n}(\omega_{k}-v_{F}k_{n})}
\Big)
\delta_{k+k_1+k_2}
\delta_{\omega_k+\omega_{k_1}+\omega_{k_2}}
+(\theta'\leftrightarrow\theta'')
\\
&\qquad
+5\text{-permutations}
\Bigg)
\end{aligned}
\end{equation}

\end{minipage}
};

\end{tikzpicture}
\end{center}

\vspace{0.8em}

\vspace{0.8em}

\begin{equation}
\label{eqn: WZW_Int_solid_LF}
\begin{aligned}
\vcenter{\hbox{%
\begin{tikzpicture}[line cap=round, line join=round, scale=1.0]

\tikzset{
    prop/.style={line width=1.2pt},
    dottedprop/.style={
        line width=1.1pt,
        dashed,
        dash pattern=on 4pt off 4pt
    },
    vertex/.style={circle, fill=black, inner sep=1.6pt},
    lab/.style={font=\small}
}

\coordinate (V) at (0,0);
\node[vertex] at (V) {};

\draw[prop] (V) -- (0,1.7);
\draw[prop] (V) -- (-1.25,-1.25);
\draw[dottedprop] (V) -- (1.35,-1.0);

\node[draw, circle, inner sep=1pt] at (-0.67,-0.6) {$\times$};
\node[lab] at (-0.67,-0.2) {$\mathcal{F}$};

\node[lab] at (0.55,1.00)
{$\langle\phi\phi\rangle_{0}$};

\node[lab] at (-1.75,-0.35)
{$\langle\phi\phi\rangle_{1}$};

\node[lab] at (1.95,-0.35)
{$\langle\phi\,\partial_{\theta}\,\phi\rangle_{0}$};

\end{tikzpicture}%
}}
\qquad
&=
i\frac{1}{3!}\frac{v_F}{2p_F^3}
\Bigg[
\mathcal{F}_{\theta,\theta'}
\frac{k_{n'}k_{1s'}}
     {(\omega_k-v_Fk_n)(\omega_k-v_Fk_{n'})}
\frac{\omega_k+2\omega_{k_1}}
     {k_{1n'}(\omega_{k_1}-v_Fk_{1n'})}
\\[0.6em]
&\qquad\times\partial_{\theta'}\!\left(
\frac{\delta_{\theta'',\theta'}}
     {k_{2n''}(\omega_{k_2}-v_Fk_{2n''})}
\right)
\Bigg]+5\text{-permutations}.
\end{aligned}
\end{equation}

\vspace{0.8em}

\begin{center}
\begin{tikzpicture}[line cap=round, line join=round, scale=1.0]

\tikzset{
    prop/.style={line width=1.2pt},
    dottedprop/.style={line width=1.1pt, dashed, dash pattern=on 4pt off 4pt},
    vertex/.style={circle, fill=black, inner sep=1.6pt},
    lab/.style={font=\small}
}

\coordinate (V) at (0,0);
\node[vertex] at (V) {};

\draw[prop] (V) -- (0,1.7);
\draw[prop] (V) -- (-1.25,-1.25);
\draw[dottedprop] (V) -- (1.35,-1.0);

\node[draw, circle, inner sep=1 pt] at (0.72,-0.53) {$\times$};
\node[lab] at (0.95,-0.20) {$\mathcal{F}$};

\node[lab] at (0.55,1.00) {$\langle \phi\phi\rangle_{0}$};
\node[lab] at (-1.75,-0.35) {$\langle \phi\phi\rangle_{0}$};
\node[lab] at (2.05,-0.35)
{$\langle \phi\,\partial_{\theta}\,\phi\rangle_{1}$};

\node[anchor=west, align=left] at (3.2,0.15) {%
\normalsize
\refstepcounter{equation}\label{eqn: WZW_Int_dased_LF}%
$
\begin{aligned}
=\, &i\frac{1}{3!}\frac{v_F}{2p_F^3}
\Bigg[
\frac{\delta_{\theta',\theta}\,k_{1s}}
     {(\omega_k-v_{F}k_n)}
\frac{\omega_k+2\omega_{k_1}}
     {k_{1n'}(\omega_{k_1}-v_{F}k_{1n'})}
\partial_{\theta}\!\left(
\frac{\mathcal{F}_{\theta'',\theta}}
     {(\omega_{k_2}-v_{F}k_{2n''})
      (\omega_{k_2}-v_{F}k_{2n})}
\right)
\Bigg]\notag\\
&\qquad+5\text{-permutations}
\end{aligned}
\quad
(\theequation)
$%
};

\end{tikzpicture}
\end{center}
\vspace{0.8em}
\begin{center}
\begin{tikzpicture}[line cap=round, line join=round, scale=1.0]

\tikzset{
    prop/.style={line width=1.2pt},
    dottedprop/.style={line width=1.1pt, dashed, dash pattern=on 4pt off 4pt},
    kvertex/.style={rectangle, fill=red, draw=red, inner sep=2.2pt},
    lab/.style={font=\small}
}

\coordinate (V) at (0,0);
\node[kvertex] at (V) {};

\draw[prop] (V) -- (0,1.7);
\draw[prop] (V) -- (-1.25,-1.25);
\draw[prop] (V) -- (1.35,-1.0);

\node[draw, circle, inner sep=1pt] at (0.68,-0.5) {$\times$};
\node[lab] at (0.7,-0.1) {$\mathcal{F}$};

\node[lab] at (0.55,1.00) {$\langle \phi\phi\rangle_{0}$};
\node[lab] at (-1.75,-0.35) {$\langle \phi\phi\rangle_{0}$};
\node[lab] at (1.95,-0.35) {$\langle \phi\,\phi\rangle_{1}$};

\node[anchor=west, align=left] at (3.2,0.15) {%
\normalsize
\refstepcounter{equation}\label{eqn: epsilon_DP_diagram}%
$
\begin{aligned}
=\, &-\frac{i}{3!}\frac{v_F}{2p_F^3}
\left(\frac{v_F}{2}+p_F\epsilon''\right)
\mathcal{F}_{\theta,\theta'}
\frac{k_{2n'}(k_1+k_2)_{n'}}
     {(\omega_k-v_{F}k_n)(\omega_k-v_{F}k_{n'})(\omega_{k_1}-v_{F}k_{1n'})}
\\[0.4em]
&\qquad \times
\frac{\delta_{\theta'',\theta'}}
     {k_{2n''}(\omega_{k_2}-v_{F}k_{2n''})}
(2\pi)^3
\delta_{k+k_1+k_2}\,
\delta_{\omega_k+\omega_{k_1}+\omega_{k_2}} +5\text{-permutations}
\end{aligned}
\qquad
(\theequation)
$%
};
\end{tikzpicture}
\end{center}
Now we evaluate the diagrams corresponding to the second-order density correction, 
which give rise to the second term in Eq.~\eqref{eqn: three-point density correlation expansion} of the form $\langle\phi^{\theta}_{\mathbf{k}}\phi^{\theta'}_{\mathbf{k}_1}
\partial_{\theta''}\phi^{\theta''}_{\mathbf{p}}\phi^{\theta''}_{\mathbf{k}_2-\mathbf{p}}\rangle$. 
\begin{align}
    \label{B16}
\langle\phi^{\theta}_{\mathbf{k}}\phi^{\theta'}_{\mathbf{k}_1}
\partial_{\theta''}\phi^{\theta''}_{\mathbf{p}}\phi^{\theta''}_{\mathbf{k}_2-\mathbf{p}}\rangle &= \langle\phi^{\theta}_{\mathbf{k}}\phi^{\theta'}_{\mathbf{k}_1}
\partial_{\theta''}\phi^{\theta''}_{\mathbf{p}}\phi^{\theta''}_{\mathbf{k}_2-\mathbf{p}} \exp[-i(S_{int})]\rangle^{\text{connected}}_{0}\notag\\
&= \langle\phi^{\theta}_{\mathbf{k}}\phi^{\theta'}_{\mathbf{k}_1}
\partial_{\theta''}\phi^{\theta''}_{\mathbf{p}}\phi^{\theta''}_{\mathbf{k}_2-\mathbf{p}}\rangle^{\text{connected}}_{0} -i\langle\phi^{\theta}_{\mathbf{k}}\phi^{\theta'}_{\mathbf{k}_1}
\partial_{\theta''}\phi^{\theta''}_{\mathbf{p}}\phi^{\theta''}_{\mathbf{k}_2-\mathbf{p}}S^{[2]}_{H,\mathcal{F}^{(2,0)}}\rangle^{\text{connected}}_{0}+\mathcal{O}(\mathcal{F}^2)+\cdots
\end{align}

\begin{center}
\begin{tikzpicture}[line cap=round,line join=round,scale=1.0]

\tikzset{
    prop/.style={line width=1.15pt},
    dprop/.style={line width=1.05pt,dashed,dash pattern=on 4pt off 4pt},
    vertex/.style={circle,fill=black,inner sep=1.6pt},
    ins/.style={circle,draw=black,line width=0.9pt,inner sep=1.2pt},
    lab/.style={font=\small},
    biglab/.style={font=\small},
    brace/.style={decorate,decoration={brace,amplitude=6pt}}
}


\coordinate (A) at (-6.5,0.0);
\node[vertex] at (A) {};

\draw[prop] (A) -- ++(0,0.5);
\draw[prop] (A) -- ++(-1.15,-1.75);
\draw[prop] ($(A)+(-0.08,-0.05)$) -- ++(-1.15,-1.75);
\draw[prop] (A) -- ++(1.25,-1.75);

\node[biglab] at (-3.9,-0.35) {$=$};


\coordinate (B) at (-1.9,0.0);
\node[vertex] at (B) {};

\draw[prop] (B) -- ++(0,0.5);
\draw[prop] (B) -- ++(-1.15,-1.75);
\draw[dprop] (B) -- ++(1.15,-1.75);

\node[lab] at (-3.2,-0.85) {$\langle \phi\phi\rangle_0$};
\node[lab] at (-0.65,-0.85) {$\langle \phi\,\partial_\theta\phi\rangle_0$};

\draw[brace] (-2.85,1.62) -- (-0.95,1.62)
node[midway,yshift=0.42cm] {$\mathcal{O}(1)$};

\draw[brace,decoration={brace,mirror,amplitude=6pt}]
(-3.15,-2.25) -- (-0.65,-2.25)
node[midway,yshift=-0.55cm] {$\rho^{(2)}$ Free part};

\node[biglab] at (0.75,-0.35) {$+$};


\coordinate (C) at (2.55,0.0);
\node[vertex] at (C) {};

\draw[prop] (C) -- ++(0,0.5);
\draw[prop] (C) -- ++(-1.15,-1.75);
\draw[dprop] (C) -- ++(1.15,-1.75);

\coordinate (XC) at ($(C)!0.48!($(C)+(1.15,-1.75)$)$);
\node[ins] at (XC) {$\times$};
\node[lab] at ($(XC)+(0.16,-0.6)$) {$\mathcal{F}$};

\node[lab] at (1.25,-0.85) {$\langle \phi\phi\rangle_0$};
\node[lab] at (3.95,-0.65) {$\langle \phi\,\partial_\theta\phi\rangle_1$};

\node[biglab] at (5.15,-0.35) {$+$};


\coordinate (D) at (6.85,0.0);
\node[vertex] at (D) {};

\draw[prop] (D) -- ++(0,0.5);
\draw[prop] (D) -- ++(-1.15,-1.75);
\draw[dprop] (D) -- ++(1.15,-1.75);

\coordinate (XD) at ($(D)!0.48!($(D)+(-1.15,-1.75)$)$);
\node[ins] at (XD) {$\times$};
\node[lab] at ($(XD)+(0.2,-0.4)$) {$\mathcal{F}$};

\node[lab] at (5.58,-0.8) {$\langle \phi\phi\rangle_1$};
\node[lab] at (8.05,-0.65) {$\langle \phi\,\partial_\theta\phi\rangle_0$};

\draw[brace] (1.30,1.62) -- (8.20,1.62)
node[midway,yshift=0.42cm] {$\mathcal{O}(\mathcal{F})$};

\draw[brace,decoration={brace,mirror,amplitude=6pt}]
(1.10,-2.25) -- (8.30,-2.25)
node[midway,yshift=-0.58cm] {$\rho^{(2)}- S_H^{(2)}\ \text{part}$};
\node[biglab] at (9.5,-0.35) {$+\cdots$};
\end{tikzpicture}
\end{center}
Let us first clarify the diagrammatic notation used above. The vertex appearing in 
$\langle 
\phi^{\theta}_{\mathbf{k}}
\phi^{\theta'}_{\mathbf{k}_1}
\partial_{\theta''}\phi^{\theta''}_{\mathbf{p}}
\phi^{\theta''}_{\mathbf{k}_2-\mathbf{p}}
\rangle$ 
is not the same as the cubic interaction vertices considered earlier. For example, in a term such as 
$\langle 
\phi^{\theta}_{\mathbf{k}}
\phi^{\theta'}_{\mathbf{k}_1}
\phi^{\theta''}_{\mathbf{k}_2}
S^{[3]}_{\mathrm{WZW}}
\rangle$, 
the fields coming from $S^{[3]}_{\mathrm{WZW}}$ form an internal interaction vertex, which is then contracted with the three external fields. In the present case, however, the factor 
$\partial_{\theta''}\phi^{\theta''}_{\mathbf{p}}
\phi^{\theta''}_{\mathbf{k}_2-\mathbf{p}}$ 
comes from the second-order density correction itself and already carries the external momentum $\mathbf{k}_2$. Therefore, we represent it as a composite density vertex attached directly to the external $\mathbf{k}_2$ leg. This is why the diagram is drawn as a wedge/triangle with a short line at the top, rather than as an ordinary cubic interaction vertex. As before, solid lines represent the free propagator 
$\langle \phi\phi\rangle_{0}$, whereas dashed lines represent the mixed propagator 
$\langle \phi\,\partial_{\theta}\phi\rangle_{0}$. The first-order correction from the Landau interaction parameter $\mathcal{F}$ is denoted by a cross insertion on a propagator. Since this insertion may be placed either on the solid line or on the dashed line, the $\mathcal{O}(\mathcal{F})$ contribution from the second-order density correction consists of two diagrams.\\
\par\noindent

We now write the explicit forms of these diagrams in momentum space:

\vspace{0.8em}

\begin{center}
\begin{tikzpicture}[line cap=round, line join=round, scale=1.0]

\tikzset{
    prop/.style={line width=1.2pt},
    dprop/.style={line width=1.1pt, dashed, dash pattern=on 4pt off 4pt},
    vertex/.style={circle, fill=black, inner sep=1.6pt},
    lab/.style={font=\small},
    smalllab/.style={font=\small},
    brace/.style={decorate, decoration={brace, amplitude=6pt}}
}

\coordinate (B) at (0,0);
\node[vertex] at (B) {};

\draw[prop] (B) -- ++(0,0.55);
\draw[prop] (B) -- ++(-1.2,-1.75);
\draw[dprop] (B) -- ++(1.2,-1.75);

\node[smalllab] at (-1.95,-0.85) {$\langle \phi\phi\rangle_0$};
\node[smalllab] at (1.95,-0.85) {$\langle \phi\,\partial_\theta\phi\rangle_0$};

\node[anchor=west, align=left] at (3.0,-1.0) {%
\normalsize
\refstepcounter{equation}\label{eqn: rho_2_free}%
$
\begin{aligned}
=\, &-\frac{1}{p_F^2}\,
\partial_{\theta''}\!\left(
\frac{\delta_{\theta,\theta''}}
{k_n(\omega_k-v_{F}k_n)}
\right)
(2\pi)^3\delta_{k+p}\,
\delta_{\omega_k+\omega_p}
\\[0.5em]
&\quad\times
\frac{\delta_{\theta',\theta''}}
{k_{1n}(\omega_{k_1}-v_{F}k_{1n})}
(2\pi)^3\delta_{k_1+k_2-p}\,
\delta_{\omega_{k_1}+\omega_{k_2}-\omega_p}
\\[0.5em]
&\quad +(k\leftrightarrow k_1)
\end{aligned}
\qquad
(\theequation)
$%
};

\end{tikzpicture}
\end{center}
\vspace{0.8em}

\begin{center}
\begin{tikzpicture}[line cap=round, line join=round, scale=1.0]

\tikzset{
    prop/.style={line width=1.2pt},
    dprop/.style={line width=1.1pt, dashed, dash pattern=on 4pt off 4pt},
    vertex/.style={circle, fill=black, inner sep=1.6pt},
    ins/.style={circle, draw=black, line width=0.8pt, inner sep=1pt},
    lab/.style={font=\small},
    smalllab/.style={font=\small},
    brace/.style={decorate, decoration={brace, amplitude=6pt}}
}

\coordinate (D) at (0,0);
\node[vertex] at (D) {};

\draw[prop] (D) -- ++(0,0.55);
\draw[prop] (D) -- ++(-1.2,-1.75);
\draw[dprop] (D) -- ++(1.2,-1.75);

\coordinate (XD) at ($(D)!0.48!($(D)+(-1.2,-1.75)$)$);
\node[ins] at (XD) {$\times$};
\node[smalllab] at ($(XD)+(0.45,-0.05)$) {$\mathcal{F}$};

\node[smalllab] at (-2.0,-0.8) {$\langle \phi\phi\rangle_1$};
\node[smalllab] at (2.0,-0.8) {$\langle \phi\,\partial_\theta\phi\rangle_0$};

\node[anchor=west, align=left] at (3.0,-1.0) {%
\normalsize
\refstepcounter{equation}\label{eqn: rho_2_Solid_V}%
$
\begin{aligned}
=\, &\frac{v_F}{2p_F^2}\,
\mathcal{F}_{\theta'',\theta}\,
\partial_{\theta''}\!\left(
\frac{\delta_{\theta',\theta''}}
{k_{1n'}(\omega_{k_1}-v_{F}k_{1n'})}
\right)
(2\pi)^3\delta_{k_1+p}\,
\delta_{\omega_{k_1}+\omega_p}
\\[0.5em]
&\quad\times
\frac{1}{(\omega_k-v_{F}k_n)}\,
\frac{1}{(\omega_k-v_{F}k_{n''})}\,
(2\pi)^3\delta_{k+k_2-p}\,
\delta_{\omega_{k_2}+\omega_k-\omega_p}
\\[0.5em]
&\quad +(k\leftrightarrow k_1)
\end{aligned}
\qquad
(\theequation)
$%
};
\end{tikzpicture}
\end{center}
\vspace{0.8em}

\begin{center}
\begin{tikzpicture}[line cap=round, line join=round, scale=1.0]

\tikzset{
    prop/.style={line width=1.2pt},
    dprop/.style={line width=1.1pt, dashed, dash pattern=on 4pt off 4pt},
    vertex/.style={circle, fill=black, inner sep=1.6pt},
    ins/.style={circle, draw=black, line width=0.8pt, inner sep=1pt},
    lab/.style={font=\small},
    smalllab/.style={font=\small},
    brace/.style={decorate, decoration={brace, amplitude=6pt}}
}

\coordinate (C) at (0,0);
\node[vertex] at (C) {};

\draw[prop]  (C) -- ++(0,0.55);
\draw[prop]  (C) -- ++(-1.2,-1.75);
\draw[dprop] (C) -- ++(1.2,-1.75);

\coordinate (XC) at ($(C)!0.48!($(C)+(1.2,-1.75)$)$);
\node[ins] at (XC) {$\times$};
\node[smalllab] at ($(XC)+(0.45,-0.05)$) {$\mathcal{F}$};

\node[smalllab] at (-2.0,-0.85) {$\langle \phi\phi\rangle_0$};
\node[smalllab] at (2.2,-0.70) {$\langle \phi\,\partial_\theta\phi\rangle_1$};

\node[anchor=west, align=left] at (3.2,-1) {%
\normalsize
\refstepcounter{equation}\label{eqn: rho_2_dased_V}%
$
\begin{aligned}
=\, &\frac{v_F}{2p_F^2}\,
\frac{\delta_{\theta,\theta''}}
{k_n(\omega_k-v_{F}k_n)(\omega_{k_1}-v_{F}k_{1n'})}
(2\pi)^3
\delta_{k_1+p}\,
\delta_{\omega_p+\omega_{k_1}}
\\[0.5em]
&\quad\times
\partial_{\theta''}\!\left(
\frac{\mathcal{F}_{\theta'\theta''}\,k_{1n''}}
{p_{n''}(\omega_p-v_{F}p_{n''})}
\right)
(2\pi)^3
\delta_{k+k_2-p}\,
\delta_{\omega_{k_2}+\omega_k-\omega_p}
\\[0.5em]
&\quad +(k\leftrightarrow k_1)
\end{aligned}
\qquad
(\theequation)
$%
};

\end{tikzpicture}
\end{center}

\subsection{Derivation of the Noninteracting Equal-Time Three-Point Density Correlator}
\label{app: Derivation_NI_correlations}
Three Feynman diagrams contribute to the noninteracting equal-time three-point density correlation:
\begin{center}
\begin{tikzpicture}[scale=0.77, line cap=round, line join=round]

\tikzset{
  lvertex/.style={circle, fill=black, inner sep=1.6pt},
  kvertex/.style={rectangle, fill=red, draw=red, inner sep=2.2pt},
  lab/.style={font=\small},
  smalllab/.style={font=\footnotesize},
  prop/.style={thick},
  dprop/.style={thick, dashed}
}

\node at (-7.8,1.0)
{$\left\langle \delta\rho_k\,\delta\rho_{k_1}\,\delta\rho_{k_2}\right\rangle_{\mathrm{NI}}$};

\node at (-4.3,1.0) {$=$};

\coordinate (V1) at (-1.5,0.8);

\node[lvertex] at (V1) {};

\draw[prop]  (V1) -- ++(0,2.0);
\draw[prop]  (V1) -- ++(-1.3,-1.5);
\draw[dprop] (V1) -- ++( 1.4,-1.5);

\node[smalllab] at ($(V1)+(0.95,1.55)$) {$\langle \phi\phi\rangle_0$};
\node[smalllab] at ($(V1)+(-1.55,-0.55)$) {$\langle \phi\phi\rangle_0$};
\node[smalllab] at ($(V1)+(1.95,-0.55)$) {$\langle \phi\,\partial_\theta\phi\rangle_0$};

\node at (1.4,1.0) {$+$};

\coordinate (V2) at (4.0,0.8);

\node[kvertex] at (V2) {};

\draw[prop] (V2) -- ++(0,2.0);
\draw[prop] (V2) -- ++(-1.3,-1.5);
\draw[prop] (V2) -- ++( 1.4,-1.5);

\node[smalllab] at ($(V2)+(0.95,1.55)$) {$\langle \phi\phi\rangle_0$};
\node[smalllab] at ($(V2)+(-1.55,-0.55)$) {$\langle \phi\phi\rangle_0$};
\node[smalllab] at ($(V2)+(1.75,-0.55)$) {$\langle \phi\phi\rangle_0$};

\node at (6.8,1.0) {$+$};

\coordinate (V3) at (9.6,0.8);
\node[lvertex] at (V3) {};

\draw[prop]  (V3) -- ++(0,0.55);
\draw[prop]  (V3) -- ++(-1.2,-1.75);
\draw[dprop] (V3) -- ++( 1.2,-1.75);

\node[smalllab] at ($(V3)+(-1.75,-0.85)$) {$\langle \phi\phi\rangle_0$};
\node[smalllab] at ($(V3)+(1.95,-0.85)$) {$\langle \phi\,\partial_\theta\phi\rangle_0$};
\end{tikzpicture}
\end{center}

The explicit form of the three-point density-density correlation function in momentum space is obtained by substituting Eqs.~\eqref{eqn: WZW free diagram}, \eqref{eqn: H3_free}, and \eqref{eqn: rho_2_free} into Eq.~\eqref{eqn: three-point density correlation expansion}. The final expression is:
\begin{align}
\label{B20}
\langle\delta\rho_{\mathbf{k}}\delta\rho_{\mathbf{k}_1}
\delta\rho_{\mathbf{-k-k_1}}\rangle_{NI}
&=
\langle\delta\rho_{\mathbf{k}}\delta\rho_{\mathbf{k}_1}
\delta\rho_{\mathbf{-k-k_1}}\rangle^{S^{[3]}_{WZW}}_{NI}
+\langle\delta\rho_{\mathbf{k}}\delta\rho_{\mathbf{k}_1}
\delta\rho_{\mathbf{-k-k_1}}\rangle^{S_H^{[3]}}_{NI}
+\langle\delta\rho_{\mathbf{k}}\delta\rho_{\mathbf{k}_1}
\delta\rho_{\mathbf{-k-k_1}}\rangle^{\delta\rho^{[2]}}_{NI}
\notag\\
&=\frac{1}{3!(2\pi)^3}\int_{\theta}
\frac{k_{n}}{(\omega_{k}-v_{F}k_{n})}
\frac{k_{1s}}{(\omega_{k1}-v_{F}k_{1n})}
\partial_{\theta}\Big(\frac{\omega_{k}+2\omega_{k_1}}{(\omega_{k}+\omega_{k_1}-v_{F}(k_{n}+k_{1n})}\Big)
+ (5-\text{permutations})
\notag\\
&\quad -\frac{1}{(2\pi)^3}\int_{\theta}
\Big(\frac{v_{F}}{2}+p_{F}\epsilon''\Big)
\frac{k_{n}}{(\omega_{k}-v_{F}k_{n})}
\frac{k_{1n}}{(\omega_{k_1}-v_{F}k_{1n})}
\frac{(k+k_1)_{n}}{(\omega_{k}+\omega_{k_1}-v_{F}(k_{n}+k_{1n}))}
\notag\\
&\quad -\frac{1}{2(2\pi)^3}\int_{\theta}
\frac{k_{n}(k+k_1)_{s}}{\omega_{k}-v_{F}k_{n}}
\partial_{\theta}\Big(\frac{1}{(\omega_{k_1}-v_{F}k_{1n})}\Big)
+(5-\text{permutations})
\end{align}
For the equal-time three-point density-density correlation function, we integrate over the frequencies:
\begin{align}
\label{B21}
\int_{\omega_{\textbf{k}},\omega_{\textbf{k}_1}}
\langle \delta \rho_{\mathbf{k}}\delta\rho_{\mathbf{k_1}}\delta\rho_{\mathbf{-k-k_1}}\rangle_{NI}
&= \int_{\omega_{\textbf{k}},\omega_{\textbf{k}_1}}
\Big(
\langle\delta\rho_{\mathbf{k}}\delta\rho_{\mathbf{k_1}}\delta\rho_{\mathbf{-k-k_1}}\rangle^{S_H^{[3]}}_{NI}
+\langle\delta\rho_{\mathbf{k}}\delta\rho_{\mathbf{k_1}}\delta\rho_{\mathbf{-k-k_1}}\rangle^{S^{[3]}_{WZW}}_{NI}
+\langle\delta\rho_{\mathbf{k}}\delta\rho_{\mathbf{k_1}}\delta\rho_{\mathbf{-k-k_1}}\rangle^{\delta\rho^{[2]}}_{NI}
\Big)
\end{align}
We evaluate the three contributions in turn:
\begin{align}
\label{B22}
\int_{\omega_{\textbf{k}},\omega_{\textbf{k}_1}}
\langle\delta\rho_{\mathbf{k}}\delta\rho_{\mathbf{k_1}}\delta\rho_{\mathbf{-k_1-k}}\rangle^{S_H^{(3)}}_{NI}
&= -\frac{1}{(2\pi)^3}
\int_{\omega_{\textbf{k}},\omega_{\textbf{k}_1},\theta}
\Big(\frac{v_{F}}{2}+p_{F}\epsilon''\Big)
\frac{k_{n}}{(\omega_{k}-v_{F}k_{n})}
\frac{k_{1n}}{(\omega_{k_1}-v_{F}k_{1n})}
\notag\\
&\qquad
\frac{(k+k_1)_{n}}{(\omega_{k}+\omega_{k_1}-v_{F}(k_{n}+k_{1n}))}
\end{align}
Let us first perform the $\omega_{\mathbf{k}}$ integral. The poles in the complex $\omega_{\mathbf{k}}$ plane are $\omega_{\mathbf{k}}=v_F k_n-i\epsilon\,\mathrm{sgn}(v_F k_n)$ and $\omega_{\mathbf{k}}=-\omega_{\mathbf{k}_1}+v_F(k+k_1)_n-i\epsilon\,\mathrm{sgn}\!\left(v_F(k+k_1)_n\right)$. If both poles lie in the same half-plane, either the upper or lower half-plane, then the contour can be closed in the opposite half-plane and the integral vanishes. Therefore, a nonzero contribution can arise only when the two poles lie on opposite sides of the real axis. This requires $k_n>0$ and $(k+k_1)_n<0$, or alternatively $k_n<0$ and $(k+k_1)_n>0$. After performing the \(\omega_{\mathbf{k}}\) contour integral, the remaining
\(\omega_{\mathbf{k}_1}\) integral contains a double pole which vanishes upon $\omega_{k_1}$ integration:
\begin{align}
\label{B23}
\int_{\omega_{\textbf{k}},\omega_{\textbf{k}_1}}\langle\delta\rho_{\mathbf{k}}\delta\rho_{\mathbf{k_1}}\delta\rho_{\mathbf{-k_1-k}}\rangle^{S_H^{(3)}}_{NI} & = -\frac{i}{(2\pi)^2}\big(\frac{v_{F}}{2}+p_{F}\epsilon''\big) \int_{\omega_{\textbf{k}_1},\theta}\frac{k_{n}k_{1n}(k+k_1)_{n}}{(\omega_{\textbf{k}_1}-v_{F}k_{1n}\pm i\epsilon)^2}=0
\end{align}
 One might also worry about the special case in which $k_n=0$ while $k_{1n}\neq 0$, or vice versa, since the Feynman pole prescription then crosses the origin. In the present integral, however, the numerator vanishes in this limit, so no additional contribution arises. This subtlety can nevertheless be important in other frequency integrals, as discussed below. Therefore, the \(S_H^{(3)}\) term does not
contribute to the noninteracting equal-time three-point density correlation.
\par\noindent

We next evaluate the $S^{(3)}_{WZW}$ contribution to the density-density correlation:
\begin{align}
\label{B24}
\int_{\omega_{k},\omega_{k_1}}
\langle \delta\rho_{k} \delta\rho_{k_1} \delta\rho_{-k-k_1}\rangle^{S^{(3)}_{WZW}}_{NI}
&=\frac{1}{3! (2\pi)^2}
\int_{\theta,\omega_{k},\omega_{k_1}}
\frac{k_{n}}{(\omega_{k}-v_{F}k_{n})}
\frac{k_{1s}}{(\omega_{k_1}-v_{F}k_{1n})}
\notag\\
&\qquad
\partial_{\theta}\Big(\frac{\omega_{k}+2\omega_{k_1}}{(\omega_{k}+\omega_{k_1})-v_{F}(k+k_1)_{n}}\Big)
+5-\text{permutations}
\notag\\
&=\frac{1}{3!(2\pi)^2}\Bigg(
\int_{\theta,\omega_{k},\omega_{k_1}}
\frac{k_{n}}{(\omega_{k}-v_{F}k_{n})}
\frac{k_{1s}}{(\omega_{k_1}-v_{F}k_{1n})}
\notag\\
&\qquad\quad
\frac{v_{F}(\omega_{k}+2\omega_{k_1})(k+k_1)_{s}}{\big((\omega_{k}+\omega_{k_1})-v_{F}(k+k_1)_{n}\big)^2}
\notag\\
&\qquad -\int_{\theta,\omega_{k},\omega_{k_1}}
\frac{(\omega_{k}+2\omega_{k_1})k_{n}k_{1s}}{(\omega_{k}-v_{F}k_{n})(\omega_{k_1}-v_{F}k_{1n})}
\frac{v_{F}k_{2s}}{(\omega_{k_2}-v_{F}k_{2n})^2}
\notag\\
&\qquad -\int_{\theta,\omega_{k},\omega_{k_1}}
\frac{(\omega_{k_1}+2\omega_{k})k_{1n}k_{s}}{(\omega_{k_1}-v_{F}k_{1n})(\omega_{k}-v_{F}k_{n})}
\frac{v_{F}k_{2s}}{(\omega_{k_2}-v_{F}k_{2n})^2}
\notag\\
&\qquad -2\int_{\theta,\omega_{k},\omega_{k_1}}
\frac{(\omega_{k_2}+2\omega_{k_1})k_{2n}k_{1s}}{(\omega_{k_2}-v_{F}k_{2n})(\omega_{k_1}-v_{F}k_{1n})}
\frac{v_{F}k_{s}}{(\omega_{k}-v_{F}k_{n})^2}
\notag\\
&\qquad -2\int_{\theta,\omega_{k},\omega_{k_1}}
\frac{(\omega_{k}+2\omega_{k_2})k_{n}k_{2s}}{(\omega_{k_2}-v_{F}k_{2n})(\omega-v_{F}k_{n})}
\frac{v_{F}k_{1s}}{(\omega_{k_1}-v_{F}k_{1n})^2}
\Bigg)
\end{align}

where $k_{2}=-k-k_{1}$ and $\omega_{k_{2}}=-\omega_k-\omega_{k_1}$. Notice that the first two integrals have the same pole structure, so it is sufficient to analyze the first one:
\begin{align}
\label{B25}
&\int_{\theta,\omega_{k},\omega_{k_1}}
\frac{(\omega_{k}+2\omega_{k_1})k_{n}k_{1s}}
{(\omega_{k}-v_{F}k_{n})(\omega_{k_1}-v_{F}k_{1n})}
\frac{(k+k_1)_{s}}
{(\omega_{k}+\omega_{k_1}-v_{F}(k+k_1)_{n})^2}
= 2\pi i 
\int_{\theta,\omega_{k_1}}
\frac{(2\omega_{k_1}-v_{F}k_{n})k_{n}k_{1s}(k+k_1)_{s}}
{(\omega_{k_1}-v_{F}k_{1n}+i\epsilon)^3}
=0 .
\end{align}
In deriving the second line, we assume $k_n\neq 0$ and $k_{1n}\neq 0$. We then perform the $\omega_k$ integral by taking the two $\omega_k$ poles to lie on opposite sides of the contour. This leaves a cubic pole in $\omega_{k_1}$. Since the numerator is at most linear in $\omega_{k_1}$, the corresponding residue vanishes. One might worry about the special case $k_{1n}=0$ with $k_n\neq 0$, where the pole prescription appears to cross the origin. The converse case, $k_n=0$ with $k_{1n}\neq 0$, is harmless because the integrand already vanishes due to the explicit factor of $k_n$ in the numerator. For $k_{1n}=0$, however, the pole prescription ensures that no additional contribution arises. In this limit, the relevant singular factors take the schematic form $(\omega_k-k_n+i\epsilon\,\mathrm{sgn}(k_n))(\omega_k+\omega_{k_1}+i\epsilon\,\mathrm{sgn}(k_n))$. Thus the two $\omega_k$ poles lie on the same side of the contour. The $\omega_k$ contour can therefore be closed on the opposite side, and the contribution vanishes. Now we study the last two terms of Eq.~\eqref{B24}. These two terms have the same pole structure, differing only by the exchange
$\omega_k\leftrightarrow \omega_{k_1}$ and $k_n\leftrightarrow k_{1n}$. Therefore, it is sufficient to analyze one of them:
\begin{align}
\label{B26}
&\int_{\theta,\omega_{k},\omega_{k_1}}
\frac{(\omega_{k_2}+2\omega_{k_1})k_{2n}k_{1s}k_{s}}
{(\omega_{k}+\omega_{k_1}-v_{F}(k+k_1)_{n})(\omega_{k_1}-v_{F}k_{1n})(\omega_{k}-v_{F}k_{n})^2}
= 2\pi i
\int_{\theta,\omega_{k}}
\frac{(-\omega_{k}+v_{F}k_{1n})k_{2n}k_{1s}k_{s}}
{(\omega_{k}-v_{F}k_{n}+i\epsilon)^3}
=0 .
\end{align}
In the above equation, we first assume $k_n\neq 0$ and $k_{1n}\neq 0$. We then perform the $\omega_{k_1}$ integral by taking the two $\omega_{k_1}$ poles to lie on opposite sides of the contour. This leaves a cubic pole in the remaining $\omega_k$ integral. Since the numerator is at most linear in $\omega_k$, the corresponding residue vanishes.

One might also worry about possible contributions from the special regions $k_n=0$, $k_{1n}=0$, or both. If both $k_n=0$ and $k_{1n}=0$, the contribution vanishes trivially. If instead only one of them vanishes, for example $k_n=0$ with $k_{1n}\neq 0$, the pole structure is again harmless. The relevant denominators take the schematic form
$
\bigl(\omega_k+\omega_{k_1}-k_{1n}
+i\epsilon\,\mathrm{sgn}(k_{1n})\bigr)
\bigl(\omega_{k_1}-k_{1n}
+i\epsilon\,\mathrm{sgn}(k_{1n})\bigr).
$
Thus the two $\omega_{k_1}$ poles lie on the same side of the contour, so the $\omega_{k_1}$ integral vanishes. The case $k_{1n}=0$ with $k_n\neq 0$ is identical after exchanging
$k\leftrightarrow k_1$. Combining these observations, the total equal-time contribution from the noninteracting WZW cubic term vanishes:
\begin{align}
\label{B27}
\int_{\omega_{k},\omega_{k_1}}
\left\langle
\delta\rho_{k}\,\delta\rho_{k_1}\,\delta\rho_{-k-k_1}
\right\rangle^{S^{(3)}_{\mathrm{WZW}}}_{\mathrm{NI}}
=0 .
\end{align}
We next evaluate the triangle/wedge diagram corresponding to $\rho^{(2)}$:
\begin{align}
\label{B28}
\int_{\omega_{k},\omega_{k_1}}
\langle \delta \rho_{k},\delta\rho_{k_1}\,\delta\rho_{k_2}\rangle^{\rho^{(2)}}_{NI}
&= -\frac{1}{2(2\pi)^3}
\int_{\theta,\omega_{k},\omega_{k_1}}
\frac{k_{n}(k+k_1)_{s}}{\omega_{k}-v_{F}k_{n}}
\partial_{\theta}\Big(\frac{1}{(\omega_{k_1}-v_{F}k_{1n})}\Big)
+(5-\text{permutations})
\notag\\
&=-\frac{1}{2(2\pi)^3}
\int_{\theta,\omega_{k},\omega_{k_1}}
\frac{k_{n}(k+k_1)_{s}}{\omega_{k}-v_{F}k_{n}}
\frac{v_{F}k_{1s}}{(\omega_{k_1}-v_{F}k_{1n})^2} +(5-\text{permutations})
\notag\\
&=-\frac{1}{2(2\pi)^3}
\int_{\theta,\omega_{k},\omega_{k_1}}
\frac{k_{n}(k+k_1)_{s}}{(\omega_{k}-v_{F}k_{n})}
\frac{v_{F}k_{1s}}{(\omega_{k_1}-v_{F}k_{1n})^2}
\notag\\
&\quad -\frac{1}{2(2\pi)^3}
\int_{\theta,\omega_{k},\omega_{k_1}}
\frac{k_{1n}(k+k_1)_{s}}{(\omega_{k_1}-v_{F}k_{1n})}
\frac{v_{F}k_{s}}{(\omega_{k}-v_{F}k_{n})^2}
\notag\\
&\quad -\frac{1}{2(2\pi)^3}
\int_{\theta,\omega_{k},\omega_{k_1}}
\frac{k_{n}(k+k_2)_{s}}{(\omega_{k}-v_{F}k_{n})}
\frac{v_{F}k_{2s}}{(\omega_{k_2}-v_{F}k_{2n})^2}
\notag\\
&\quad -\frac{1}{2(2\pi)^3}
\int_{\theta,\omega_{k},\omega_{k_1}}
\frac{k_{2n}(k_2+k_1)_{s}}{(\omega_{k_2}-v_{F}k_{2n})}
\frac{v_{F}k_{1s}}{(\omega_{k_1}-v_{F}k_{1n})^2}
\notag\\
&\quad -\frac{1}{2(2\pi)^3}
\int_{\theta,\omega_{k},\omega_{k_1}}
\frac{k_{1n}(k_2+k_1)_{s}}{(\omega_{k_1}-v_{F}k_{1n})}
\frac{v_{F}k_{2s}}{(\omega_{k_2}-v_{F}k_{2n})^2}
\notag\\
&\quad -\frac{1}{2(2\pi)^3}
\int_{\theta,\omega_{k},\omega_{k_1}}
\frac{k_{2n}(k+k_2)_{s}}{(\omega_{k_2}-v_{F}k_{2n})}
\frac{v_{F}k_{s}}{(\omega_{k}-v_{F}k_{n})^2}
\end{align}
We now evaluate the remaining frequency integrals using the Sokhotski--Plemelj formula,
\[
\frac{1}{\omega_k-v_F k_n+i\epsilon\,\mathrm{sgn}(\omega_k)}
=
\mathcal{P}\!\left(\frac{1}{\omega_k-v_F k_n}\right)
-i\pi\,\mathrm{sgn}(\omega_k)\,
\delta(\omega_k-v_F k_n),
\]
together with
\[
\int_{\omega_k}
\frac{d\omega_k}
{\bigl(\omega_k-v_F k_n+i\epsilon\,\mathrm{sgn}(\omega_k)\bigr)^2}
=
-\frac{2i\pi}{v_{F}}\,\delta(k_n),
\]
Note that, unlike in the previous integrals involving $S_{\mathrm{WZW}}$ and $S^{(3)}_{H}$, the special cases $k_n\neq 0$ with $k_{1n}=0$, or vice versa, are important here because of the delta-function contribution from the Sokhotski-Plemelj formula. Applying these identities gives
\begin{align*}
 \int_{\omega_{k},\omega_{k_1}}
\left\langle 
\delta \rho_{k}\,\delta\rho_{k_1}\,\delta\rho_{k_2}
\right\rangle^{\rho^{(2)}}_{\mathrm{NI}}
&=
\frac{1}{(2\pi)^2}
\Bigg[
|\vec k\times \vec k_1|
+\frac{1}{2}
\bigl(|k|^2+|k_1|^2\bigr)
|\sin(\varphi_1-\varphi)|
\cos(\varphi_1-\varphi)
\notag\\
&\hspace{5.2cm}
-\frac{1}{2}
\bigl(|k|^2+|k_1|^2\bigr)
|\sin(\varphi_1-\varphi)|
\cos(\varphi_1-\varphi)
\Bigg].   
\end{align*}
\begin{align}
\label{B29}
    \int_{\omega_{k},\omega_{k_1}}
\left\langle 
\delta \rho_{k}\,\delta\rho_{k_1}\,\delta\rho_{-k-k_1}
\right\rangle^{\rho^{(2)}}_{\mathrm{NI}}
=\frac{1}{(2\pi)^2}|\vec k\times \vec k_1| .
\end{align}
Here $\varphi_1-\varphi$ denotes the angle between $\vec k_1$ and $\vec k$. The final answer is therefore written purely in terms of the magnitude of the two-dimensional cross product between the external momenta. Combining the contributions in Eqs.~\eqref{B23}, \eqref{B27}, and \eqref{B29}, we obtain,
\begin{align}
\label{B30}
\int_{\omega_{k},\omega_{k_1}}
\left\langle 
\delta \rho_{\mathbf{k}}\,
\delta\rho_{\mathbf{k}_1}\,
\delta\rho_{\mathbf{-k-k_1}}
\right\rangle_{NI}
=
\frac{1}{(2\pi)^2}
|\mathbf{k}\times\mathbf{k}_1| .
\end{align}

\subsection{The \texorpdfstring{$\epsilon''(p_F)$}{epsilon''(pF)} Contribution to the Interacting Density Three-Point Function}
\label{app: epsillon_dprine_derivation}
In the interacting case, the analysis becomes considerably more involved, with six diagrams contributing at order $\mathcal{O}(\mathcal{F})$: Eqs.~\eqref{eqn: WZW_Int_vertex}-\eqref{eqn: epsilon_DP_diagram}, \eqref{eqn: rho_2_Solid_V}, and \eqref{eqn: rho_2_dased_V}. Since the contour integrations are already delicate in the noninteracting theory, we isolate the diagram proportional to the band curvature $\epsilon''$. This contribution is particularly useful because it is independent of the other diagrams at linear order in the interaction and can therefore be evaluated separately.
\begin{equation}
\label{B31}
\begin{tikzpicture}[baseline=(eq.base), line cap=round, line join=round, scale=0.9]

\tikzset{
    prop/.style={line width=1.2pt},
    kvertex/.style={rectangle, fill=red, draw=red, inner sep=2.2pt},
    lab/.style={font=\small}
}

\node[anchor=base west, inner sep=0pt] (eq) at (0,0) {$
\displaystyle
\left\langle
\delta \rho_{\mathbf{k}}\,
\delta \rho_{\mathbf{k}_1}\,
\delta \rho_{-\mathbf{k}-\mathbf{k}_1}
\right\rangle_{\mathcal{O}(\mathcal{F})}^{\epsilon^{\prime\prime}}
=
$};

\begin{scope}[shift={(7.8,0)}]

\coordinate (V) at (0,0);
\node[kvertex] at (V) {};

\draw[prop] (V) -- (0,1.6);
\draw[prop] (V) -- (-1.25,-1.25);
\draw[prop] (V) -- (1.35,-1.25);

\node[draw, circle, inner sep=1pt] at (0.68,-0.65) {$\times$};
\node[lab] at (0.82,-0.12) {$\mathcal{F}$};

\node[lab] at (0.0,1.9) {$\mathbf{k}$};
\node[lab] at (-1.55,-1.43) {$\mathbf{k}_1$};
\node[lab] at (1.75,-1.4) {$\mathbf{k_2}=-\mathbf{k}-\mathbf{k}_1$};

\end{scope}
\end{tikzpicture}
\end{equation}
To obtain the explicit momentum-space expression, we substitute Eq.~\eqref{eqn: epsilon_DP_diagram} into Eq.~\eqref{eqn: three-point density correlation expansion} and use the symmetry of the Landau interaction,
$\mathcal{F}_{\theta,\theta'}=\mathcal{F}_{\theta',\theta}$ which gives:
\begin{align}
\label{B32}
    \langle
\delta \rho_{\mathbf{k}}\,
\delta \rho_{\mathbf{k}_1}\,
\delta \rho_{\mathbf{k_2}}
\rangle_{\mathcal{O}(\mathcal{F})}^{\epsilon^{\prime\prime}}
&=-\frac{ v_{F}p_{F}}{6(2\pi)^3}\bigg[\int_{\theta,\theta'}\epsilon''\mathcal{F}^{(2,0)}_{\theta,\theta'}\frac{k_{2n}k_{2n'}k_{1n'}k_{n'}}{(\omega_{k_2}-v_{F}k_{2n})(\omega_{k_2}-v_{F}k_{2n'})(\omega_{k}-v_{F}k_{n'})(\omega_{k_1}-v_{F}k_{1n'})}\notag\\
&\hspace{12 em}+\int_{\theta,\theta'}\epsilon''\mathcal{F}^{(2,0)}_{\theta,\theta'}\frac{k_{2n}k_{n}k_{1n'}k_{1n}}{(\omega_{k_2}-v_{F}k_{2n})(\omega_{k_1}-v_{F}k_{1n})(\omega_{k_1}-v_{F}k_{1n'})(\omega_{k}-v_{F}k_{n})}
\notag\\
&\hspace{12 em}+\int_{\theta,\theta'}\epsilon''\mathcal{F}^{(2,0)}_{\theta,\theta'}\frac{k_{2n}k_{n}k_{1n}k_{n'}}{(\omega_{k_2}-v_{F}k_{2n})(\omega_{k}-v_{F}k_{n})(\omega_{k}-v_{F}k_{n'})(\omega_{k_1}-v_{F}k_{1n})}\bigg]
\end{align}
where $\mathbf{k}_2=-\mathbf{k}-\mathbf{k}_1$ and
$\omega_{\mathbf{k}_2}=-\omega_{\mathbf{k}}-\omega_{\mathbf{k}_1}$.
To obtain the equal-time three-point density-density correlation function, we integrate over the frequencies. We perform this integration explicitly for the three terms above.

\begin{align}
\label{B33} \int_{\omega_{k},\omega_{k_1}}\text{Term}\,\mathrm{I}  
&= -\frac{ v_{F}p_{F}\epsilon''(p_{F})}{6(2\pi)^3}\int_{\theta,\theta',\omega_{k},\omega_{k_1}}\frac{\mathcal{F}_{\theta,\theta'}(k+k_1)_{n}(k+k_1)_{n'}}{(\omega_{k}+\omega_{k_1}-v_{F}(k+k_1)_{n})(\omega_{k}+\omega_{k_1}-v_{F}(k+k_1)_{n'})}\notag\\
&\hspace{16 em}\times\frac{k_{1n'}k_{n'}}{(\omega_{k}-v_{F}k_{n'})(\omega_{k_1}-v_{F}k_{1n'})}
\end{align}

We first perform the $\omega_k$ integral. The poles in the $\omega_k$ plane are located at 
$\omega_k=-\omega_{k_1}+v_F(k+k_1)_n-i\epsilon\,\mathrm{sgn}\!\big((k+k_1)_n\big)$, 
$\omega_k=-\omega_{k_1}+v_F(k+k_1)_{n'}-i\epsilon\,\mathrm{sgn}\!\big((k+k_1)_{n'}\big)$, 
and $\omega_k=v_F k_{n'}-i\epsilon\,\mathrm{sgn}(k_{n'})$. 
For the contour integral to give a nonzero contribution, the poles must be split between the upper and lower half-planes; equivalently, two poles must lie in one half of the complex plane, while the remaining pole lies in the opposite half-plane.  Moreover, one can readily check that the only nonzero contributions arise when 
$(k+k_1)_n>0$, $(k+k_1)_{n'}<0$, and $k_{n'}<0$, 
or, alternatively, when 
$(k+k_1)_n<0$, $(k+k_1)_{n'}>0$, and $k_{n'}>0$. 
After carrying out the $\omega_k$ integral, the remaining $\omega_{k_1}$ integral reduces to a straightforward simple-pole contour integral, with the sign fixed by the two cases $k_{1n'}<0$ and $k_{1n'}>0$, respectively. It is also useful to examine the potentially subtle limits in which one of the normal momentum components vanishes. As discussed in the previous section, the limits $k_{n'}=0$ with $k_{1n'}\neq 0$, and vice versa, require special care because the pole prescription changes as the pole crosses the real axis. In the present contribution, however, no additional singular term is generated. For example, when $k_{n'}=0$ and $k_{1n'}\neq 0$, the first and third terms in Eq.~\eqref{B32} vanish explicitly because their numerators are proportional to $k_{n'}$. The second term remains finite and has a well-defined pole structure. Therefore, this limit does not produce any extra singular contribution to the frequency integral.  Combining the two nonzero sign configurations, we obtain
\begin{small}
\begin{align}
\label{B34}
\int_{\omega_{k},\omega_{k_1}}\mathrm{Term}\,\mathrm{I}
&=  
-\frac{ p_{F}\epsilon''}{6(2\pi)v_{F}}
|k_1|\,|k|
\int_{\theta,\theta'}
\mathcal{F}^{(2,0)}_{\theta,\theta'}
\frac{
k_{2n}k_{2n'}
(\hat{k}_{1}\cdot\hat{n}')(\hat{k}\cdot\hat{n}')
}{
\left[k_{2n}-k_{2n'}\right]^2
}
\notag\\
&\qquad\times
\bigg[
\Theta\!\left(-\mathrm{sgn}\!\left(k_{2n}\right)\right)
\Theta\!\left(-\mathrm{sgn}\!\left((k_1)_{n'}\right)\right)
\Theta\!\left(-\mathrm{sgn}\!\left(k_{n'}\right)\right)
+
\Theta\!\left(\mathrm{sgn}\!\left(k_{2n}\right)\right)
\Theta\!\left(\mathrm{sgn}\!\left((k_1)_{n'}\right)\right)
\Theta\!\left(\mathrm{sgn}\!\left(k_{n'}\right)\right)
\bigg] .
\end{align}
\end{small}
where again $k_2=-k-k_1$. The angular constraints ensure that 
$\mathrm{sgn}\!\left(k_{2n}\right)$ and 
$\mathrm{sgn}\!\left(k_{2n'}\right)$ are opposite. 
Therefore, the denominator 
$\left[k_{2n}-k_{2n'}\right]^2$ never becomes singular in the contributing region. By a similar argument, the remaining two integrals are
\begin{small}
\begin{align}
\label{B35}
\int_{\omega_{k},\omega_{k_1}}\mathrm{Term}\,\mathrm{II}
&=  -\frac{ p_{F}\epsilon''}{6(2\pi)v_{F}}
|k||k_2|\int_{\theta,\theta'}\frac{\mathcal{F}^{(2,0)}_{\theta,\theta'}(\hat{k}.\hat{n})(\hat{k_2}.\hat{n})k_{1n}k_{1n'}}{(k_{1n'}-k_{1n})^2}\notag\\&\qquad\quad\times\bigg[\Theta(-\text{sgn}(k_{2n}))\Theta(-\text{sgn}(k_{n}))\Theta(-\text{sgn}(k_{1n'}))+\Theta(\text{sgn}(k_{2n}))\Theta(\text{sgn}(k_{n}))\Theta(\text{sgn}(k_{1n'}))\bigg]
\end{align}
\end{small}
\begin{small}
    \begin{align}
       \label{B36}
\int_{\omega_{k},\omega_{k_1}}\mathrm{Term}\,\mathrm{III}
&=-\frac{ p_{F}\epsilon''}{6(2\pi)v_{F}}
|k_{1}||k_2|\int_{\theta,\theta'}\frac{\mathcal{F}^{(2,0)}_{\theta,\theta'}k_{n}k_{n'}(\hat{k_1}.\hat{n})(\hat{k_2}.\hat{n})}{(k_{n'}-k_{n})^2}\notag\\
&\qquad\quad\times\bigg[\Theta(-\text{sgn}(k_{2n}))\Theta(-\text{sgn}(k_{1n}))\Theta(-\text{sgn}(k_{n'}))+\Theta(\text{sgn}(k_{2n}))\Theta(\text{sgn}(k_{1n}))\Theta(\text{sgn}(k_{n'})\bigg]
    \end{align}
\end{small}
Collecting all contributions, we obtain the $\epsilon''$-dependent interaction correction to the equal-time three-point density-density correlation function.
\begin{align}
    \label{B38}
     \int_{\omega_{k},\omega_{k_1}}\langle
\delta \rho_{\mathbf{k}}\,
\delta \rho_{\mathbf{k}_1}\,
\delta \rho_{\mathbf{k_2}}
\rangle_{\mathcal{O}(\mathcal{F})}^{\epsilon^{\prime\prime}}&=\int_{\omega_{k},\omega_{k_1}}\Big(\mathrm{Term}\,\mathrm{I}+\mathrm{Term}\,\mathrm{II}+\mathrm{Term}\,\mathrm{III}\Big)\notag\\
&=-\frac{ p_{F}\epsilon''}{12\pi v_{F}}
|k_1|\,|k|
\int_{\theta,\theta'}
\mathcal{F}^{(2,0)}_{\theta,\theta'}
\frac{
k_{2n}k_{2n'}
(\hat{k}_{1}\cdot\hat{n'})(\hat{k}\cdot\hat{n'})
}{
\left[k_{2n}-k_{2n'}\right]^2
}
\notag\\
&\qquad\times
\bigg[
\Theta\!\left(-\mathrm{sgn}\!\left(k_{2n}\right)\right)
\Theta\!\left(-\mathrm{sgn}\!\left((k_1)_{n'}\right)\right)
\Theta\!\left(-\mathrm{sgn}\!\left(k_{n'}\right)\right)
+
\Theta\!\left(\mathrm{sgn}\!\left(k_{2n}\right)\right)
\Theta\!\left(\mathrm{sgn}\!\left((k_1)_{n'}\right)\right)
\Theta\!\left(\mathrm{sgn}\!\left(k_{n'}\right)\right)
\bigg]\notag\\
&\qquad-\frac{ p_{F}\epsilon''}{12\pi v_{F}}|k||k_2|\int_{\theta,\theta'}\frac{\mathcal{F}^{(2,0)}_{\theta,\theta'}(\hat{k}.\hat{n})(\hat{k_2}.\hat{n})k_{1n}k_{1n'}}{(k_{1n'}-k_{1n})^2}\notag\\&\qquad\quad\times\bigg[\Theta(-\text{sgn}(k_{2n}))\Theta(-\text{sgn}(k_{n}))\Theta(-\text{sgn}(k_{1n'}))+\Theta(\text{sgn}(k_{2n}))\Theta(\text{sgn}(k_{n}))\Theta(\text{sgn}(k_{1n'}))\bigg]\notag\\
&\qquad-\frac{ p_{F}\epsilon''}{12\pi v_{F}}|k_{1}||k_2|\int_{\theta,\theta'}\frac{\mathcal{F}^{(2,0)}_{\theta,\theta'}k_{n}k_{n'}(\hat{k_1}.\hat{n})(\hat{k_2}.\hat{n})}{(k_{n'}-k_{n})^2}\notag\\
&\qquad\quad\times\bigg[\Theta(-\text{sgn}(k_{2n}))\Theta(-\text{sgn}(k_{1n}))\Theta(-\text{sgn}(k_{n'}))+\Theta(\text{sgn}(k_{2n}))\Theta(\text{sgn}(k_{1n}))\Theta(\text{sgn}(k_{n'})\bigg]
&\qquad
\end{align}
 To show explicitly that the above integral is nonzero, we consider the simplest case in which the Landau parameter is constant,
$\mathcal{F}^{(2,0)}_{\theta,\theta'}=\mathcal{F}_{0}$, and take $|k|=|k_1|$. In this case, we obtain
\begin{align}
\label{B39}
\int_{\omega_k,\omega_{k_1}}
\left\langle
\delta \rho_{\mathbf{k}}\,
\delta \rho_{\mathbf{k}_1}\,
\delta \rho_{\mathbf{k}_2}
\right\rangle_{\mathcal{O}(\mathcal{F})}^{\epsilon''}
&=
-\frac{ p_F \epsilon''}{12\pi v_F}
|k|^2 \mathcal{F}_0
\notag\\
&\qquad\times
\int_{\theta,\theta'}
\frac{
\big[\cos(\theta-\varphi)+\cos(\theta-\varphi_1)\big]
\big[\cos(\theta'-\varphi)+\cos(\theta'-\varphi_1)\big]
\cos(\theta'-\varphi)\cos(\theta'-\varphi_1)
}{
\big[
\cos(\theta-\varphi)+\cos(\theta-\varphi_1)
-\cos(\theta'-\varphi)-\cos(\theta'-\varphi_1)
\big]^2
}
\notag\\
&\hspace{4em}\times
\Big[
\Theta\!\left(\text{sgn}\{\cos(\theta-\varphi)+\cos(\theta-\varphi_1)\}\right)
\Theta\!\left(\text{sgn}\{-\cos(\theta'-\varphi_1)\}\right)
\Theta\!\left(\text{sgn}\{-\cos(\theta'-\varphi)\}\right)
\notag\\
&\hspace{6em}
+
\Theta\!\left(\text{sgn}\{-\cos(\theta-\varphi)-\cos(\theta-\varphi_1)\}\right)
\Theta\!\left(\text{sgn}\{\cos(\theta'-\varphi_1)\}\right)
\Theta\!\left(\text{sgn}\{\cos(\theta'-\varphi)\}\right)
\Big]
\notag\\
&\qquad
+\frac{p_F \epsilon''}{12\pi v_F}
|k|^2 \mathcal{F}_0
\int_{\theta,\theta'}
\frac{
\big[\cos(\theta-\varphi)+\cos(\theta-\varphi_1)\big]
\cos(\theta-\varphi)\cos(\theta-\varphi_1)
\cos(\theta'-\varphi_1)
}{
\big[\cos(\theta'-\varphi_1)-\cos(\theta-\varphi_1)\big]^2
}
\notag\\
&\hspace{4em}\times
\Big[
\Theta\!\left(\text{sgn}\{\cos(\theta-\varphi)+\cos(\theta-\varphi_1)\}\right)
\Theta\!\left(\text{sgn}\{-\cos(\theta-\varphi)\}\right)
\Theta\!\left(\text{sgn}\{-\cos(\theta'-\varphi_1)\}\right)
\notag\\
&\hspace{6em}
+
\Theta\!\left(\text{sgn}\{-\cos(\theta-\varphi)-\cos(\theta-\varphi_1)\}\right)
\Theta\!\left(\text{sgn}\{\cos(\theta-\varphi)\}\right)
\Theta\!\left(\text{sgn}\{\cos(\theta'-\varphi_1)\}\right)
\Big]
\notag\\
&\qquad
+\frac{p_F \epsilon''}{12\pi v_F}
|k|^2 \mathcal{F}_0
\int_{\theta,\theta'}
\frac{
\big[\cos(\theta-\varphi)+\cos(\theta-\varphi_1)\big]
\cos(\theta-\varphi)\cos(\theta-\varphi_1)
\cos(\theta'-\varphi)
}{
\big[\cos(\theta'-\varphi)-\cos(\theta-\varphi)\big]^2
}
\notag\\
&\hspace{4em}\times
\Big[
\Theta\!\left(\text{sgn}\{\cos(\theta-\varphi)+\cos(\theta-\varphi_1)\}\right)
\Theta\!\left(\text{sgn}\{-\cos(\theta-\varphi_1)\}\right)
\Theta\!\left(\text{sgn}\{-\cos(\theta'-\varphi)\}\right)
\notag\\
&\hspace{6em}
+
\Theta\!\left(\text{sgn}\{-\cos(\theta-\varphi)-\cos(\theta-\varphi_1)\}\right)
\Theta\!\left(\text{sgn}\{\cos(\theta-\varphi_1)\}\right)
\Theta\!\left(\text{sgn}\{\cos(\theta'-\varphi)\}\right)
\Big]
\notag\\
&=
-\frac{p_F \epsilon''}{12\pi v_F}
|k|^2 \mathcal{F}_0\,\mathcal{I}(\varphi -\varphi_1).
\end{align}

\subsection{Nonanalyticity near \texorpdfstring{$\Delta\varphi\rightarrow\pi$}{Delta phi -> pi}}
\label{app: cusp}
In this section, we demonstrate explicitly that the nonanalytic behavior near $\Delta\varphi=\pi$ is not a numerical artifact, but an intrinsic feature of the
interaction correction. As in the previous section, we restrict to
$|\mathbf{k}|=|\mathbf{k}_1|\equiv k$ and take the Landau interaction to be angle
independent, $\mathcal{F}^{(2,0)}_{\theta,\theta'}=\mathcal{F}_0$. It is
convenient to exploit rotational invariance and rotate the coordinate system by
the average angle
\begin{equation}
    \bar{\varphi}=\frac{\varphi+\varphi_1}{2},
\end{equation}
after which the two external directions sit symmetrically about the $x$-axis,
\begin{equation}
    \varphi\rightarrow\varphi-\bar{\varphi}=\frac{\Delta\varphi}{2},
    \qquad
    \varphi_1\rightarrow\varphi_1-\bar{\varphi}=-\frac{\Delta\varphi}{2},
\end{equation}
with $\Delta\varphi\equiv\varphi-\varphi_1$. We approach the antiparallel point
by setting
\begin{equation}
    \Delta\varphi=\pi+\delta,
    \qquad |\delta|=|\pi-\Delta\varphi|\ll 1,
\end{equation}
where the two signs of $\delta$ correspond to the two sides of the cusp. For
$\hat{\mathbf{n}}_{\theta}=(\cos\theta,\sin\theta)$ the normal projections are
\begin{align}
    k_{n}&=k\sin\Big(\theta-\tfrac{\delta}{2}\Big)=k\sin\theta+\mathcal{O}(\delta),\notag\\
    k_{1n}&=-k\sin\Big(\theta+\tfrac{\delta}{2}\Big)=-k\sin\theta+\mathcal{O}(\delta),\notag\\
    k_{2n}&=2k\sin\tfrac{\delta}{2}\,\cos\theta=k\delta\cos\theta+\mathcal{O}(\delta^{3}),
\end{align}
Analogous expressions hold for projections onto
$\hat{\mathbf n}_{\theta'}=(\cos\theta',\sin\theta')$. We now evaluate Eq.~\eqref{B38} term by term.

\paragraph{Term I.} Carrying out the frequency integrals and inserting the
projections above,
\begin{align}
    \int_{\omega_{k},\omega_{k_1}}\mathrm{Term}\,\mathrm{I}
    &= -\frac{p_{F}\epsilon''}{12\pi v_{F}}k^2\mathcal{F}_{0}
    \int_{\theta,\theta'}
    \frac{\cos\theta\cos\theta'}
    {(\cos\theta-\cos\theta')^2}
    \Big(\sin^2\tfrac{\delta}{2}-\sin^2\theta'\Big)
    \notag\\
    &\qquad\times
    \Bigg(
    \Theta\big(-\mathrm{sgn}(k\delta\cos\theta)\big)
    \Theta\big(\mathrm{sgn}(\sin(\theta'+\tfrac{\delta}{2}))\big)
    \Theta\big(-\mathrm{sgn}(\sin(\theta'-\tfrac{\delta}{2}))\big)
    \notag\\
    &\qquad\quad+
    \Theta\big(\mathrm{sgn}(k\delta\cos\theta)\big)
    \Theta\big(-\mathrm{sgn}(\sin(\theta'+\tfrac{\delta}{2}))\big)
    \Theta\big(\mathrm{sgn}(\sin(\theta'-\tfrac{\delta}{2}))\big)
    \Bigg).
\end{align}
The step functions force $k_{n'}$ and $k_{1n'}$ to carry the same sign, which
near $\Delta\varphi=\pi$ reads
\begin{equation}
    k_{n'}k_{1n'}>0
    \quad\Longrightarrow\quad
    \sin^2\theta'<\sin^2\tfrac{\delta}{2}
    \quad\Longrightarrow\quad
    |\sin\theta'|<\big|\sin\tfrac{\delta}{2}\big|.
\end{equation}
For small $|\delta|$ this confines $\theta'$ to two narrow windows around
$\theta'=0$ and $\theta'=\pi$, each of width
\begin{equation}
    \Delta\theta'\sim|\delta|.
\end{equation}
Inside those windows the numerator is itself small,
$k_{n'}k_{1n'}\sim\mathcal{O}(\delta^{2})$, and the shrinking measure supplies one
further power, so that
\begin{equation}
    \mathcal{I}_{\mathrm{I}}\sim\mathcal{O}(|\delta|^{3}).
\end{equation}
Term I therefore contributes no cusp and can be dropped at leading order.

\paragraph{Terms II and III.} These two contributions map onto one another under
$\mathbf{k}\leftrightarrow\mathbf{k}_1$, so it is enough to evaluate one and
double the result. Term II carries an explicit prefactor $|\mathbf{k}_2|$, which
by the discussion above is $2k|\sin(\delta/2)|\simeq k\,|\pi-\Delta\varphi|$:
\begin{align}
  \int_{\omega_{k},\omega_{k_1}}\mathrm{Term}\,\mathrm{II}
    &= -\frac{p_{F}\epsilon''}{12\pi v_{F}}k^2\mathcal{F}_{0}\,|\pi-\Delta\varphi|
    \int_{\theta,\theta'}
    \frac{\sin^2\theta\,\cos\theta\,\sin\theta'}
    {(\sin\theta-\sin\theta')^2}
    \Big(
    \Theta(-\mathrm{sgn}(\delta\cos\theta))
    \Theta(-\mathrm{sgn}(\sin\theta))
    \Theta(\mathrm{sgn}(\sin\theta'))
    \notag\\
    &\qquad\qquad\qquad\qquad
    +\Theta(\mathrm{sgn}(\delta\cos\theta))
    \Theta(\mathrm{sgn}(\sin\theta))
    \Theta(-\mathrm{sgn}(\sin\theta'))
    \Big).
\end{align}
The remaining question is whether the angular integral survives the above limit. The
domains selected by the step functions are listed in Table~\ref{tab:termII}.

\begin{table}[h]
\centering
\begin{tabular}{c|c|c|c}
\hline
$\mathrm{sgn}(\delta)$
& branch
& $\theta$
& $\theta'$ \\
\hline
$+$ & $k_{2n},k_n,k_{1n'}<0$ & $\pi<\theta<3\pi/2$   & $0<\theta'<\pi$ \\
$+$ & $k_{2n},k_n,k_{1n'}>0$ & $0<\theta<\pi/2$      & $\pi<\theta'<2\pi$ \\
\hline
$-$ & $k_{2n},k_n,k_{1n'}<0$ & $3\pi/2<\theta<2\pi$  & $0<\theta'<\pi$ \\
$-$ & $k_{2n},k_n,k_{1n'}>0$ & $\pi/2<\theta<\pi$    & $\pi<\theta'<2\pi$ \\
\hline
\end{tabular}
\caption{Angular domains selected by the step functions in
$\mathrm{Term}\,\mathrm{II}$ near $\Delta\varphi=\pi$.}
\label{tab:termII}
\end{table}

Unlike $\mathrm{Term}\mathrm{I}$, the allowed angular regions remain of finite measure as $\Delta\varphi\rightarrow\pi$. The angular integral therefore remains $\mathcal{O}(1)$ and approaches a finite nonzero constant, which determines the coefficient $c_1$ of the leading $|\pi-\Delta\varphi|$ dependence.
\begin{equation}
    \mathcal{I}_{\mathrm{II}}=\mathcal{I}_{\mathrm{III}}
    \simeq-\tfrac{1}{2}c_{1}\,|\pi-\Delta\varphi|,
    \qquad c_{1}>0 .
\end{equation}
Adding the three contributions, Term I drops out at leading order and we obtain
\begin{equation}
    \boxed{\;
    \mathcal{I}(\Delta\varphi\rightarrow\pi)\;\approx\;-c_{1}\,|\pi-\Delta\varphi|\;}
\end{equation}
in agreement with the linear cusp found numerically in Fig.~1(c), where
$c_{1}=1.32$. Thus, the linear cusp at $\Delta\varphi=\pi$ is a genuine
nonanalytic feature of the interaction correction rather than a numerical
artifact.
\subsection{Behavior near \texorpdfstring{$\Delta\varphi\rightarrow0$}{Delta phi -> 0}}
\label{App: smooth}
In this section, we investigate the behavior near
$\Delta\varphi\rightarrow 0$, following the same approach as in the previous
section. We restrict to $|\mathbf{k}|=|\mathbf{k}_1|\equiv k$ and take the
Landau interaction to be angle independent,
$\mathcal{F}^{(2,0)}_{\theta,\theta'}=\mathcal{F}_0$. In this parallel configuration $\Delta\varphi\rightarrow 0$, the momentum projections onto $\hat{\mathbf{n}}_{\theta}=(\cos\theta,\sin\theta)$ are given by:
\begin{align}
    k_{n}&=k\cos(\theta-\delta/2) \notag\\
    k_{1n}&=k\cos(\theta+\delta/2) \notag\\
    k_{2n}&=-2k\cos{(\delta/2)}\cos{\theta}
\end{align}
We now evaluate Eq.~\eqref{B38} term by term in this $\Delta\varphi \rightarrow 0$ limit.
\paragraph{Term I.} Carrying out the frequency integrals and inserting the
projections above,
\begin{align}
    \int_{\omega_{k},\omega_{k_1}}\text{Term}\,{\mathrm{I}}&=-\frac{p_{F}\epsilon''(p_{F})}{12\pi v_{F}}k^2\mathcal{F}_{0}\int_{\theta,\theta'}\frac{\cos{\theta}\cos{\theta'}}{(\cos{\theta}-\cos{\theta'})^2}\Big(\cos^2{\theta'}-\sin^2{\delta/2}\Big)\notag\\
    &\qquad\qquad\times\Big[\Theta(\text{sgn}(\cos{\theta}))\Theta(-\text{sgn}(\cos{(\theta'-\delta/2)}))\Theta(-\text{sgn}(\cos{(\theta'+\delta/2)}))\notag\\
     &\qquad\qquad+
    \Theta(-\text{sgn}(\cos{\theta}))\Theta(\text{sgn}(\cos{(\theta'-\delta/2)}))\Theta(\text{sgn}(\cos{(\theta'+\delta/2)}))\Big]
\end{align}

The angular constraints therefore restrict the integration domain to
\begin{align}
D_{\delta}
=
\left\{
(\theta,\theta'):
\cos\theta\,\cos\theta'<0,\;
|\cos\theta'|
>
\left|\sin\frac{\delta}{2}\right|
\right\}.
\end{align}
At $\delta=0$, this reduces to
\begin{align}
D_0
=
\left\{
(\theta,\theta'):
\cos\theta\,\cos\theta'<0
\right\}.
\end{align}
Thus $D_{\delta}\subset D_0$, with
$D_0\setminus D_{\delta}$ consisting of narrow regions near
$\theta'=\pi/2$ and $3\pi/2$ whose widths scale as
$\mathcal{O}(|\delta|)$.

$\mathrm{Term}\,\mathrm{I}$ may therefore be written as
\begin{align}
\int_{\omega_k,\omega_{k_1}}
\mathrm{Term}\,\mathrm{I}
={}&
-\frac{p_F\epsilon''(p_F)}{12\pi v_F}
k^2\mathcal{F}_0
\int_{D_{\delta}}
d\theta\,d\theta'\,
\frac{\cos\theta\,\cos\theta'}
     {(\cos\theta-\cos\theta')^2}
\left[
\cos^2\theta'
-\sin^2\left(\frac{\delta}{2}\right)
\right].
\end{align}
Using
\begin{align}
\int_{D_{\delta}}
=
\int_{D_0}
-
\int_{D_0\setminus D_{\delta}},
\end{align}
we obtain
\begin{align}
\int_{\omega_k,\omega_{k_1}}
\mathrm{Term}\,\mathrm{I}
={}&
-\frac{p_F\epsilon''(p_F)}{12\pi v_F}
k^2\mathcal{F}_0
\Bigg[
\int_{D_0}
d\theta\,d\theta'\,
\frac{\cos\theta\,\cos\theta'}
     {(\cos\theta-\cos\theta')^2}
\left(
\cos^2\theta'
-\sin^2\frac{\delta}{2}
\right)
\notag\\
&\qquad\qquad
-
\int_{D_0\setminus D_{\delta}}
d\theta\,d\theta'\,
\frac{\cos\theta\,\cos\theta'}
     {(\cos\theta-\cos\theta')^2}
\left(
\cos^2\theta'
-\sin^2\frac{\delta}{2}
\right)
\Bigg].
\end{align}
The second integral is subleading in the parallel limit because
$D_0\setminus D_{\delta}$ has width $\mathcal{O}(|\delta|)$ and the
integrand vanishes sufficiently rapidly within this region. Its contribution
therefore scales as
$\mathcal{O}\!\left(\delta^4\log(1/|\delta|)\right)$.
We therefore find
\begin{align}
\int_{\omega_k,\omega_{k_1}}
\mathrm{Term}\,\mathrm{I}
={}&
-\frac{p_F\epsilon''(p_F)}{12\pi v_F}
k^2\mathcal{F}_0
\Bigg[
\int_{D_0}
d\theta\,d\theta'\,
\frac{\cos\theta\,\cos^3\theta'}
     {(\cos\theta-\cos\theta')^2}
\notag\\
&\qquad\qquad
-\frac{\delta^2}{4}
\int_{D_0}
d\theta\,d\theta'\,
\frac{\cos\theta\,\cos\theta'}
     {(\cos\theta-\cos\theta')^2}
+
\mathcal{O}\!\left(
\delta^4\log\frac{1}{|\delta|}
\right)
\Bigg].
\end{align}

The two $\delta$-independent angular integrals are finite and evaluate to
\begin{align}
\int_{D_0}
d\theta\,d\theta'\,
\frac{\cos\theta\,\cos^3\theta'}
     {(\cos\theta-\cos\theta')^2}
&=
\pi^2-12
\simeq -2.1304,
\\
\int_{D_0}
d\theta\,d\theta'\,
\frac{\cos\theta\,\cos\theta'}
     {(\cos\theta-\cos\theta')^2}
&=-4.
\end{align}
It follows that
\begin{align}
\int_{\omega_k,\omega_{k_1}}
\mathrm{Term}\,\mathrm{I}
&=
-\frac{p_F\epsilon''(p_F)}{12\pi v_F}
k^2\mathcal{F}_0
\left[
\pi^2-12+\delta^2
+
\mathcal{O}\!\left(
\delta^4\log\frac{1}{|\delta|}
\right)
\right]\notag\\
\qquad\qquad\Rightarrow\mathcal{I}_{\mathrm I}(\Delta\varphi=\delta)&=\pi^2-12+\delta^2
+
\mathcal{O}\!\left(
\delta^4\log\frac{1}{|\delta|}
\right)
\end{align}
Therefore, we already see that, in the limit $\Delta\varphi\rightarrow0$,
$\mathrm{Term}\,\mathrm{I}$ varies quadratically and remains smooth.

\paragraph{Terms II and III.}
As noted earlier, these two contributions map onto one another under
$\mathbf{k}\leftrightarrow\mathbf{k}_1$, so it is sufficient to evaluate only one of them. In the limit $\Delta \varphi \rightarrow0$, the $\mathrm{Term}\,\mathrm{II}$ integral is given by:
\begin{align}
\int_{\omega_{k},\omega_{k_1}}\text{Term}\,{\mathrm{II}}=&\frac{p_{F}\epsilon''(p_{F})}{12\pi v_{F}}k^2 \mathcal{F}_{0}\int_{\theta,\theta'}\frac{2\cos{(\delta/2)}\cos{\theta}\cos{(\theta-\delta/2)}\cos{(\theta+\delta/2)}\cos{(\theta'+\delta/2)}}{[\cos{(\theta'+\delta/2)}-\cos{(\theta+\delta/2)}]^2}\notag\\
    &\qquad\times\Big[\Theta(\text{sgn}(\cos{\theta}))\Theta (-\text{sgn}(\cos{(\theta-\delta/2)}))\Theta(-\text{sgn}(\cos{(\theta'+\delta/2)}))
    \notag\\
    &\qquad\quad+\Theta(-\text{sgn}(\cos{\theta}))\Theta (\text{sgn}(\cos{(\theta-\delta/2)}))\Theta(\text{sgn}(\cos{(\theta'+\delta/2)}))\Big]
\end{align}
The first two sign constraints restrict $\theta$ to narrow intervals of width
$\mathcal{O}(|\delta|)$ around $\pi/2$ and $3\pi/2$. Within these intervals,
the three $\theta$-dependent factors in the numerator,
$\cos(\theta-\delta/2)$, $\cos\theta$, and $\cos(\theta+\delta/2)$, are each
of order $\mathcal{O}(|\delta|)$, while the remaining $\theta'$-dependent
factor is generically finite. The numerator therefore contributes $\mathcal{O}(|\delta|^3)$. The integration
measure $d\theta$ provides an additional factor of $\mathcal{O}(|\delta|)$.
Therefore, the contributions from $\mathrm{Term}\,\mathrm{II}$ and
$\mathrm{Term}\,\mathrm{III}$ are subleading:
\begin{align}
\int_{\omega_k,\omega_{k_1}}
\mathrm{Term}\,\mathrm{II},
\quad
\int_{\omega_k,\omega_{k_1}}
\mathrm{Term}\,\mathrm{III}
\sim
\mathcal{O}(\delta^4).
\end{align}
Finally, adding all three contributions, we obtain
\begin{align}
    \mathcal{I}(\Delta\varphi=\delta)
    =
    \pi^2-12+\delta^2
    +
    \mathcal{O}\!\left(
    \delta^4\log\frac{1}{|\delta|}
    \right).
\end{align}
We therefore conclude that the $\epsilon''(p_F)$ contribution varies
quadratically near $\Delta\varphi\rightarrow 0$, remains smooth, and approaches
a finite value in the parallel limit.

\addcontentsline{toc}{section}{References}
\bibliographystyle{quantum}
\bibliography{Reference} 
\end{document}